\documentclass[preprintnumbers,article,amsmath,amssymb,floatfix,10pt,prd,onecolumn,
superscriptaddress,nofootinbib]{revtex4}
\usepackage{bbm}
\usepackage{amsfonts}
\usepackage{mathrsfs}
\usepackage{latexsym}

\usepackage{epsfig}
\usepackage{epstopdf}
\usepackage{graphicx}
\usepackage{amssymb}
\usepackage{amsmath}
\usepackage{dcolumn}
\usepackage{bm}
\usepackage{color}
\usepackage{comment}
\usepackage{xcolor}
\begin{document}

\title{\bf Physically viable solutions of anisotropic spheres in $f(\mathscr{R},\mathscr{G})$ gravity satisfying the Karmarkar condition }

\author{G. Mustafa}
\email{gmustafa3828@gmail.com}\affiliation{Department of Mathematics, Shanghai University,
Shanghai, 200444, Shanghai, People's Republic of China}
\author{M. Farasat Shamir}
\email{farasat.shamir@nu.edu.pk; farasat.shamir@gmail.com}\affiliation{National University of Computer and
Emerging Sciences,\\ Lahore Campus, Pakistan.}
\author{Xia Tie-Cheng}
\email{xiatc@shu.edu.cn}\affiliation{Department of Mathematics, Shanghai University,
Shanghai, 200444, Shanghai, People's Republic of China}

\begin{abstract}

This paper is devoted to discuss compact stars in $f(\mathscr{R},\mathscr{G})$ gravity, where $\mathscr{R}$ and $\mathscr{G}$ denote the Ricci scalar and Gauss-Bonnet invariant respectively. To meet this aim, we consider spherically symmetric space-time with anisotropic fluid distribution. In particular, the Karmarkar condition is used to explore the compact star solutions. Further, we choose two specific model of compact stars namely LMC X-4 (mass =1.29$M/M_{\odot}$ \& radii=9.711 km) and EXO 1785-248 (mass =1.30$M/M_{\odot}$ \& radii=8.849 km). We develop the field equations for $f(\mathscr{R},\mathscr{G})$ gravity by employing the Karmarkar condition with a specific model already reported in literature by Lake \cite{L}. We further consider the Schwarzschild geometry for matching conditions at the boundary. It is important to mention here that we calculate the values of all the involved parameter by imposing the matching condition. We have provided a detailed graphical analysis to discuss the physical acceptability of parameters, i.e., energy density, pressure, anisotropy, and gradients. We have also examined the stability of compacts stars by exploring the energy conditions, equation of state, generalized Tolman-Oppenheimer-Volkoff equation, causality condition, and adiabatic index. For present analysis, we predict some numerical values in tabular form for central gravitational metric functions, central density and central pressures components. We have also calculated the ratio $p_{rc}/ \rho_{c}$ to check the validity of Zeldovich's condition. Conclusively, it is found that our obtained solutions are physically viable with well-behaved nature in $f(\mathscr{R},\mathscr{G})$ modified gravity for the compact star models under discussion.
\end{abstract}

\maketitle

\date{\today}

\section{Introduction}

In modern day cosmology, compact star discussions have attracted much
attention due to their importance in understanding different astrophysical issues.
Therefore, inquiring the internal and external structures
along with physical attributes of some explicit models of compact stars become
interesting. In particular, ``compact stars'' refer to
very different objects, including neutron stars, white dwarfs, brown
dwarfs etc. We have many information about the structure of these objects, though there
are some difficulties related to some aspects of the modeling,
particularly with neutron stars.
The equilibrium of a compact star can be analyzed by employing the Tolman-Oppenheimer-Volkoff (TOV) equations in the respective theory of gravity.
In fact, TOV equation can be symbolized in terms of anisotropic, hydrostatic and gravitational forces. The balancing feature of these forces actually suggests that stellar structures under discussion are stable and physically acceptable.
Many interesting works are available in the literature on the subject.
Folomeev et al. \cite{1a} studied spherically symmetric equilibrium configurations for polytropic matter non-minimally coupled to an external chameleon scalar field. They also argued the stability analysis and it was shown that the system had static, regular and asymptotically flat nature.  Dzhunushaliev et al. \cite{2a} explored the gravitating symmetric configurations using a scalar field along with the simple stability test. Jetzer \cite{3a} investigated the dynamical stability of spherically symmetric gravitational equilibria of cold stellar objects made of bosons and fermions. In another paper, Jetzer and his collaborators \cite{4a} explored the dynamical instability of the static real scalar field using Einstein-Klein-Gordon equation.

Modified theories of gravity have been discussed with a great zeal during recent years.
The study of stellar objects in modified gravity in an intersecting topic of discussion. In particular, new theoretical stellar structures may emerge and they could have very important observational consequences with modified gravity. However, the simplest extension of general relativity (GR) namely the $f(\mathscr{R})$ gravity, some models can be rejected as they do not allow the existence of stable stellar
configurations \cite{stable1,stable2}. But stable configurations can be obtained with some suitable modified Gauss-Bonnet gravity models \cite{stable3,stable4}.  Thus $f(\mathscr{R},\mathscr{G})$ gravity due to the addition of Gauss-Bonnet term may resolve the shortcomings
of $f(\mathscr{R})$ theory, in particular, to study the stellar structures. A brief literature regarding the study of stellar structures in the context of modified gravity is as follows:
Staykov et al. \cite{0034} investigated self-consistently slowly rotating neutron and strange stars in $\mathscr{R}$-squared gravity. The numerical results showed that the neutron star moment of inertia could be up to $30\%$ larger as compared to the usual GR models.
Astashenok et al. \cite{34} explored some interesting aspects of neutron stars in the framework of modified $f(\mathscr{R})$ theory of gravity and it was showed the maximal neutron star mass satisfying the recent observational data could be achieved for a simple hyperon equations of state.
Moraes and his collaborators \cite{3400} studied the hydrostatic equilibrium configuration of neutron and strange stars using two different equations of state in $f(\mathscr{R},\mathcal{T})$ gravity with a conclusion that the maximum stellar mass matching the observational limits can be obtained for a fixed central density. In another paper \cite{3401}, charged anisotropic compact stars with a minimal geometric deformation gravitational decoupling approach have been discussed in the context of $f(\mathscr{R},\mathcal{T})$ theory. TOV equation for the energy-momentum-conserved $f(\mathscr{R},\mathcal{T})$ theory of gravity has been used to study strange quark stars with a linear equation of state and the MIT bag model \cite{3402}.
Shamir and Ahmad \cite{MeMushtaq1}-\cite{MeMushtaq3} studied the compact stars in $f(\mathscr{G},\mathcal{T})$ theory and it was shown that $f(\mathscr{G},\mathcal{T})$ gravity provides consistent results with observational data. In a recent review paper \cite{3403}, properties of stellar structure are discussed in detail in the context of modified theories of gravity.

An interesting approach for deriving the solutions of field equations in the context of compact objects, has been used by the researchers namely the Karmarkar condition.
This condition was firstly proposed by Karmarkar \cite{76} and is regarded as a compulsory condition for a spherically symmetric space-time to be of embedding class-I. It basically helps us to obtain the exact solutions of field equations.
In a recent paper \cite{14c}, we have investigated the compact stars in the background of observational data by employing the Karmarkar condition with in $f(\mathscr{G},\mathcal{T})$ gravity background and it is shown that the obtained solutions are physically arguable with well-behaved nature. Some interesting attributes of compact stars are reported with some important $f(\mathscr{R},\mathscr{G})$ gravity models \cite{14a}. It is shown that specific $f(\mathscr{R},\mathscr{G})$ gravity models in the presence of charge may provide some cosmological solutions that fit with the observational data \cite{14b}. Thus it seems interesting to further explore the modified gravity with a hope of some viable results.

In present study, we are focussed to discuss compact stars in $f(\mathscr{R},\mathscr{G})$ gravity. For this purpose, we consider spherically symmetric space-time and anisotropic source of fluid. Our main aim of this study is to explore the compact stars solutions with their necessary properties by employing the Karmarkar condition. In particular, we choose two specific model of compact stars LMC X-4 (mass =1.29$M/M_{\odot}$ \& radii=9.711 km) \cite{LMC} and EXO 1785-248 (mass =1.30$M/M_{\odot}$ \& radii=8.849 km) \cite{EXO}.
These models are important as it has been argued that in X-ray pulsars one of the most direct and reliable way to determine the magnetic field is the registration of the cyclotron absorption lines in their energy spectra \cite{Walter}. In particular, the high luminosity of both models leads to a relatively high estimate of the magnetic field on the surface of the compact star \cite{Shtykovsky, Ozel}.
A brief pattern of the paper is as follows:
Section II is used to
give some basics of $f(\mathscr{R},\mathscr{G})$ gravity with
Karmarkar condition and anisotropic matter distribution. Third section provides the matching conditions.  Section IV is dedicated for the
discussion of some physical features of the current study. Lastly, we give some
concluding remarks.

\section{$f(\mathscr{R},\mathscr{G})$ Gravity and Karmarkar condition}

The modified action for $f(\mathscr{R},\mathscr{G})$ gravity is given by
\begin{eqnarray}\label{1}
S_{Action}=\frac{1}{2\kappa}\int d^4x\sqrt{-g}f(\mathscr{R},\mathscr{G})+S_\mathscr{M}(g^{\mu\nu},\psi),\label{1}
\end{eqnarray}
where $\kappa$ is the coupling constant, $\mathscr{R}$ is Ricci scalar, $\mathscr{G}$ is Gauss-Bonnet invariant and $S_{\mathscr{M}}(g^{\mu\nu},\psi)$ is the matter action.
We get the following modified field equation for $f(\mathscr{R},\mathscr{G})$ gravity by varying action (\ref{1}) with respect to metric tensor
\begin{eqnarray}
\mathscr{R}_{\mu\nu}-\frac{1}{2}g_{\mu\nu}\mathscr{R}&=&\kappa T^{(matt)}_{\mu\nu}+\nabla_\mu\nabla_\nu f_\mathscr{R}-g_{\mu\nu}\Box f_\mathscr{R}+\mathscr{R}\nabla_\mu\nabla_\nu f_\mathscr{G}-2g_{\mu\nu}\mathscr{R}\Box f_\mathscr{G}-\nonumber \\&&4\mathscr{R}^\alpha_\mu\nabla_\alpha\nabla_\nu f_\mathscr{G}
-4\mathscr{R}^\alpha_\nu \nabla_\alpha\nabla_\mu f_\mathscr{G}+4\mathscr{R}_{\mu\nu}\Box f_\mathscr{G}+4g_{\mu\nu}\mathscr{R}^{\theta\phi}\nabla_\theta\nabla_\phi f_\mathscr{G}+\nonumber
\\&&4\mathscr{R}_{\mu\theta\phi\nu}\nabla^\theta\nabla^\phi f_\mathscr{G}-\frac{1}{2}g_{\mu\nu}V+(1-f_\mathscr{R})\mathscr{G}_{\mu\nu},\label{2}
\end{eqnarray}
where
\begin{equation*}
V\equiv f_{\mathscr{R}}\mathscr{R}+f_{\mathscr{G}}\mathscr{G}-f(\mathscr{R},\mathscr{G}),\;\;\;\;\;\;f_\mathscr{R}\equiv\frac{\partial f (\mathscr{R},\mathscr{G})}{\partial \mathscr{R}},\;\;\;\;\;\;f_\mathscr{G}\equiv\frac{\partial f (\mathscr{R},\mathscr{G})}{\partial \mathscr{G}},
\end{equation*}
and $T_{\mu\nu}^{(matt)}$ denotes the ordinary matter.
The spherically symmetric space-time is described as
\begin{eqnarray}\label{3}
ds^{2}=-(e^{b(r)}dr^{2}+r^{2}d\theta^{2}+r^{2}\sin^{2}\theta d\phi^{2})+e^{a(r)}dt^{2}.
\end{eqnarray}
The energy-momentum tensor with anisotropic fluid is given by
\begin{equation}\label{4}
T^{(matt)}_{\mu\nu}=\rho u_\alpha u_\beta+p_r v_\alpha v_\beta +p_t(u_\alpha u_\beta-g_{\alpha\beta}-v_\alpha v_\beta),
\end{equation}
where $u_\alpha=e^{a/2} \delta_\alpha^0$, $v_\alpha=e^{b/2}\delta_\alpha^1$ are four velocities. Radial and tangential pressures are $p_r$ and $p_t$ respectively, and the energy density is denoted by $\rho$.
In the current study, we consider the following $f(\mathscr{R},\mathscr{G})$ model
\begin{equation}\label{5}
f(\mathscr{R},\mathscr{G})=\mathscr{R}+\lambda \times \mathscr{R}^{2}+\mathscr{G}^2,
\end{equation}
with $\lambda$ being a constant parameter and using space-time (\ref{3}), the Ricci scalar and the Gauss-Bonnet invariant in terms of metric coefficients turn out to be
\begin{eqnarray*}
\mathscr{R}&=&\frac{1}{2} e^{-b(r)} \left(2 a''(r)-a'(r) b'(r)+a'(r)^2+\frac{4 a'(r)}{r}-\frac{4 b'(r)}{r}-\frac{4 e^{b(r)}}{r^2}+\frac{4}{r^2}\right),\\
\mathscr{G}&=&\frac{2 e^{-2 b(r)} \left(\left(e^{b(r)}-3\right) a'(r) b'(r)+\left(1-e^{b(r)}\right) \left(2 a''(r)+a'(r)^2\right)\right)}{r^2}.
\end{eqnarray*}
It is mentioned here that the chosen model (\ref{5}) includes Starobinsky like $f(\mathscr{R})$ model along with squared Gauss-Bonnet term. However, the addition of linear term in $\mathscr{G}$ makes the calculations more tedious and in present case, we are not able to find results. Moreover, the viability of this $f(\mathscr{G})$ gravity model has already been shown in cosmological contexts \cite{M1,M2}. Also, this model belongs to
the general class of the models without the irregular spin-2 ghosts \cite{M3}.

Now, we shall explore the well-known Karmarkar condition \cite{76} which is one of the most important aspect for the present study. The basic structure of Karmarkar condition depends upon the embedded Riemannian-space of class-I. Eisenhart \cite{41} calculated a necessary and sufficient condition which is based on a symmetric tensor of second order $\chi_{a\eta}$ and the Riemann curvature tensor $\mathscr{R}_{a b\eta\gamma}$,
\begin{eqnarray}\label{6}
\Sigma(\chi_{a\eta}\chi_{b\gamma}-\chi_{a\gamma}\chi_{b\eta})=\mathscr{R}_{ab\eta\gamma},~~~~~~
\chi_{ab;\eta}-\chi_{a\eta;b}=0,
\end{eqnarray}
where $\verb";"$ denotes the covariant derivative and $\Sigma=\pm 1$ corresponds to a time like or a space like manifold.
The Riemann tensor components for embedded class-1, using space-time (\ref{3}) are calculated as
\begin{eqnarray*}
\mathscr{R}_{1414}&=&\frac{e^{a (r)}(2a ''(r)+a '(r)^2-a '(r)b'(r))}{4},\;\;\;\;\;\mathscr{R}_{2323}=\frac{r^{2}sin^{2}\theta (e^{b (r)}-1)}{e^{b (r)}},\nonumber\\
\mathscr{R}_{1212}&=&\frac{rb '(r)}{2},\;\;\;\;\;\;\;\;\;\;\;\;\;\;\;\;\;\;\;\;\;\;\;\;\;\;\;\;\;\;\;\;\;\;\;\;\;\;\;\;\;\;\;\;\;\;\;\;\;\mathscr{R}_{3434}=\frac{rsin^{2}\theta b '(r) e^{a (r)-b (r)}}{2},\nonumber\\
\mathscr{R}_{1334}&=&\mathscr{R}_{1224}sin^{2}\theta,\;\;\;\;\;\;\;\;\;\;\;\;\;\;\;\;\;\;\;\;\;\;\;\;\;\;\;\;\;\;\;\;\;\;\;\;\;\;\;\;\;\;\mathscr{R}_{1224}= 0.\nonumber
\end{eqnarray*}
Now using Eq. (\ref{6}), it follows
\begin{equation}\label{8}
\mathscr{R}_{1414}\mathscr{R}_{2323}=\mathscr{R}_{1224}\mathscr{R}_{1334}+ \mathscr{R}_{1212}\mathscr{R}_{3434}.
\end{equation}
This is known as the Karmarkar condition, with a constraint $\mathscr{R}_{2323}\neq0$. The constraint $\mathscr{R}_{2323}\neq0$ denotes the Pandey Sharma condition \cite{42}. Equation (\ref{8}) provides a differential equation
\begin{equation}\label{9}
\frac{a'(r) b'(r)}{1-e^{b(r)}}-\left(a'(r) b'(r)+a'(r)^2-2 \left(a''(r)+a'(r)^2\right)\right)=0,
\end{equation}
with $e^{b(r)}\neq1$. After integration, we get a following relationship between the metric coefficients
\begin{equation}\label{10}
e^{b(r)}=e^{a(r)} a'(r)^2+C+1,
\end{equation}
where $C$ is an integration constant. Lake \cite{L} considered a specific form of a $g_{tt}$ component of the spherically symmetric space-time, which is mentioned below
\begin{equation}\label{11}
e^{a(r)}=Q \left(L r^2+1\right)^K,
\end{equation}
where $Q$, $L$ are un-known parameters, and $K$ is a positive integer greater than 2. For present analysis, we assume $K=3,5,10,20,50,100,500$. Using Eq. (\ref{11}) in Eq. (\ref{10}), we calculate the corresponding $g_{rr}$ component of the spherically symmetric space-time as
\begin{equation}\label{12}
e^{b(r)}=F\times L r^2 \left(L r^2+1\right)^{K-2}+1,
\end{equation}
where $F=4 K^{2}L Q C$. Manipulating Eqs. (\ref{3}-\ref{5}) and (\ref{11}-\ref{12}) in field equations (\ref{2}), we get the following expressions for energy density, pressure components and anisotropy distribution
\begin{eqnarray}
\rho&=&\frac{L}{r^6 \chi _5(r){}^4}\bigg(-\frac{8 K \left(L r^2-1\right) \chi _{15}(r)}{\left(L r^2+1\right)^2}+\frac{\chi _{13}(r)}{\left(L r^2+1\right)^4}-\frac{\chi _{14}(r)}{\left(L r^2+1\right)^8}+\frac{\chi _{12}(r)}{\chi _1(r){}^3}-\frac{2 F^2 L r^4 \chi _3(r){}^2}{\chi _1(r){}^2}\nonumber\\&\times&\chi _{16}(r)-\frac{\chi _{17}(r)}{\left(L r^2+1\right)^2}-\frac{2 K^3 L^2 r^5 \chi _{18}(r)}{\left(L r^2+1\right)^3}-\frac{r^2 \chi _{20}(r)}{\left(L r^2+1\right)^7}2 F r^2 +\frac{1}{\chi _1(r)} \chi _3(r)\times\bigg(-\frac{\chi _{21}(r)}{\left(L r^2+1\right)^2}\nonumber\\&+&r \bigg(\frac{\chi _{23}(r)}{\left(L r^2+1\right)^3}+\chi _{22}(r)\bigg)\bigg)+\frac{2 K^2 L r^2}{\left(L r^2+1\right)^2}\times\bigg(\frac{256 F^3 L^3 r^6 \chi _3(r){}^3 \left(-\chi _5(r)\right)}{\chi _1(r){}^3}-\frac{\chi _{25}(r)}{\left(L r^2+1\right)^6}\nonumber\\&+&\frac{\chi _{24}(r)}{\chi _1(r)}+4 \chi _{27}(r)-\frac{\chi _{26}(r)}{\chi _1(r){}^2}\bigg)-\frac{2 K L r^2}{L r^2+1}\times\bigg(\frac{16 F^2 L r^2 \chi _3(r){}^2 \chi _{29}(r)}{\chi _1(r){}^2}-\frac{F \chi _3(r) \chi _{34}(r)}{\chi _1(r)}\nonumber\\&+&\frac{4 L r^2 \chi _{33}(r)}{\left(L r^2+1\right)^7}-\frac{\chi _{31}(r)}{\left(L r^2+1\right)^6}+\frac{\chi _{28}(r)}{\chi _1(r){}^4}+\chi _{30}(r)-\frac{\chi _{32}(r)}{\chi _1(r){}^3}\bigg)\bigg),\label{13}\\
p_{r}&=&\frac{L}{\left(L r^2+1\right)^8 \chi _5(r){}^4}\times\bigg(\frac{\chi _{47}(r) \chi _{48}(r)}{\chi _{35}(r){}^3}-\frac{\chi _{46}(r)}{\chi _{35}(r){}^2}-\frac{8 \lambda  L \left(L r^2+1\right) \chi _{38}(r) \chi _{49}(r)}{\chi _{35}(r)}+\chi _{50}(r) \chi _{51}(r)\nonumber\\&-&\left(L r^2+1\right) \chi _3(r) \chi _{52}(r) \left(-2 F \left(L r^2+1\right)^K+K L r^2+K\right)\bigg),\label{14}\\
p_{t}&=&\frac{L}{\left(L r^2+1\right)^9 \chi _5(r){}^4}\bigg(-\frac{576 F^2 K^2 L^3 \chi _3(r){}^2 \left(L r^2+1\right)^{2 K+5}}{\chi _{35}(r){}^2}-\frac{128 F K L^3 \chi _{76}(r) \left(L r^2+1\right)^K}{\chi _{35}(r){}^3}\nonumber\\&-&\frac{\chi _{77}(r)}{\chi _{35}(r){}^4}-\chi _{69}(r)+\frac{8 \lambda  L \left(L r^2+1\right) \chi _{78}(r)}{\chi _{35}(r){}^2}+\left(L r^2+1\right) \chi _{50}(r) \chi _{74}(r)-\left(L r^2+1\right) \chi _{79}(r) \chi _{75}(r)\bigg),\label{15}\\
\triangle&=&\frac{L r^4}{\left(L r^2+1\right)^9 \chi _2(r){}^4}\bigg(-\frac{384 F K L^3 \chi _{80}(r) \chi _{81}(r) \left(L r^2+1\right)^{K+3}}{\chi _{35}(r){}^3}-\frac{128 F K L^3 \chi _{76}(r) \left(L r^2+1\right)^K}{\chi _{35}(r){}^3}\nonumber\\&-&\chi _{69}(r)-\frac{\chi _{77}(r)}{\chi _{35}(r){}^4}+\frac{8 \lambda  L \left(L r^2+1\right)^2 \chi _{38}(r) \chi _{82}(r)}{\chi _{35}(r)}\lambda  +\frac{1}{\chi _{35}(r){}^2} \chi _{90}(r)+\left(L r^2+1\right)^2 \chi _3(r) \chi _{88}(r) \nonumber\\&\times&\left(-2 F \left(L r^2+1\right)^K+K L r^2+K\right)-\left(L r^2+1\right) \chi _{79}(r) \chi _{89}(r)\bigg),\label{16}
\end{eqnarray}
where the computation of $\chi _i(r)$, $\{i=1,...,90\}$ is cumbersome and the final
result is far from illuminating, so we do not include it here.
\section{Matching conditions}
\begin{center}
\begin{table}
\caption{\label{tab1}{Approximated values of $K,\; Q,\;L,\;F$ and $\lambda$.}}
\begin{tabular}{|c|c|c|c|c|c|c|c|c|}
    \hline
    & \multicolumn{4}{|c|}{LMC X-4} \\
    \hline
K \;\;\;\;                               & $Q$\;\;\;\                  &$L$\;\;\;               &$F$\;\;\;\;\;\;            &$ \lambda$ \\
\hline
3\;\;\;\;                                &0.432949436117\;\;\;\;       &0.0012738366711745\;\;\;\;  &4.78207627799156\;\;\;\;   &-0.00012729167661809\;\;\;\; \\
\hline
5\;\;\;\;                                &0.436329253451\;\;\;\;       &0.0007292603756168\;\;\;\;  &7.66407984623659\;\;\;\;   &-0.00016249223959843\;\;\;\;\\
\hline
10\;\;\;\;                               &0.438748172452\;\;\;\;       &0.0003525088500154\;\;\;\;  &14.9007596461127\;\;\;\;   &-0.00023152070615084\;\;\;\;\\
\hline
20\;\;\;\;                               &0.439922722178\;\;\;\;       &0.0001733727253591\;\;\;\;  &29.3927503380711\;\;\;\;   &-0.00032060580232051\;\;\;\;\\
\hline
50\;\;\;\;                               &0.440616724155\;\;\;\;       &0.0000686753985575\;\;\;\;  &72.8828318491836\;\;\;\;   &-0.00043979823132680\;\;\;\;\\
\hline
100\;\;\;\;                              &0.440846305436\;\;\;\;       &0.0000342268670036\;\;\;\;  &145.370879184403\;\;\;\;   &-0.00050837551295063\;\;\;\;\\
\hline
500\;\;\;\;                 &0.441029346097\;\;\;\;       &   $6.82774299900\times10^{-6}$\;\;\;\;  &725.283382067673\;\;\;\;   &-0.00058419953372838\;\;\;\;\\
\hline
& \multicolumn{4}{|c|}{EXO 1785-248} \\
    \hline
3\;\;\;\;                                &0.376886170583\;\;\;\;       &0.0018624730435760\;\;\;\;  &4.56985954394590\;\;\;\;   &-0.00020162688698357\;\;\;\; \\
\hline
5\;\;\;\;                                &0.381136135653\;\;\;\;       &0.0010558873370445\;\;\;\;  &7.27776221229879\;\;\;\;   &-0.00026012665905526\;\;\;\;\\
\hline
10\;\;\;\;                               &0.384153811848\;\;\;\;       &0.0005069846571835\;\;\;\;  &14.0883492289874\;\;\;\;   &-0.00038556962325101\;\;\;\;\\
\hline
20\;\;\;\;                               &0.385612104007\;\;\;\;       &0.0002485585248070\;\;\;\;  &27.7331663550059\;\;\;\;   &-0.00057206110796352\;\;\;\;\\
\hline
50\;\;\;\;                               &0.386471652079\;\;\;\;       &0.0000982757462016\;\;\;\;  &68.6853550511597\;\;\;\;   &-0.00087766555492779\;\;\;\;\\
\hline
100\;\;\;\;                              &0.386755655891\;\;\;\;       &0.0000489495284643\;\;\;\;  &136.944732244935\;\;\;\;   &-0.00109192244615192\;\;\;\;\\
\hline
500\;\;\;\;                 &0.386981965562\;\;\;\;       &   $9.75997782892\times10^{-6}$\;\;\;\;    &683.029892944439\;\;\;\;   &-0.00137247339688812\;\;\;\;\\
\hline
\end{tabular}
\end{table}
\end{center}
In the theory of GR, the Schwarzschild's solution is considered as an appropriate choice to choose from the diverse possibilities of the matching conditions while exploring the compact stellar objects. Also according to the Jebsen-Birkhoff's theorem statement, every spherically symmetric vacuum solution of field equations must be static and asymptotically flat. Furthermore, as a concern with the modified $f(\mathscr{R},\mathscr{G})$ gravity, the Schwarzschild's solution may be accommodated with a proper choice of viable $f(\mathscr{R},\mathscr{G})$ gravity models for non zero energy density and pressure. Perhaps, this fact leads to the violation of Birkhoff's theorem in modified theories of gravity \cite{Faraoni}. In fact while studying the junction conditions for $f(\mathscr{R})$ gravity, Senovilla \cite{Senovilla} obtained the field equations for the energy-momentum tensor on the shell/brane and remarkably they turned out to be same as those in GR. Assuming similar results hold in the case of $f(\mathscr{R},\mathscr{G})$ gravity, we can join the internal geometry given by Eq. (\ref{3}) with exterior Schwarzschild space-time
\begin{equation}\label{17}
 d{s}^2=\left(1-\frac{2 M}{r}\right) dt^2-\left(1-\frac{2 M}{r}\right)^{-1} d{r}^2-r^2(d\theta^{2}+sin^{2}\theta d\phi^{2}),
\end{equation}
where, $M$ represents the mass of the stellar object.
Now, imposing metric coefficients continuity on the boundary $r=R$, we get the following constraints
\begin{eqnarray}
Q \left(L R^2+1\right)^K&=&1-\frac{2 M}{R},\label{18}\\
F L R^2 \left(L R^2+1\right)^{K-2}+1&=&\left(1-\frac{2 M}{R}\right)^{-1},\label{19}\\
K Q L \left(L R^2+1\right)^{K-1}&=& \frac{M}{R^3},\label{20}\\
p_{r}(r=R)&=&0.\label{21}
\end{eqnarray}
Utilizing these boundaries conditions from Eqs. (\ref{18}-\ref{21}), we get the following relations for the following unknowns:
\begin{eqnarray}
Q&=&\frac{(R-2 M) \left(1-\frac{M}{2 K M-K R+M}\right)^{-K}}{r},\label{22}\\
L&=&\frac{M}{R^2 (K R-(2 K+1) M)},\label{23}\\
F&=&2 K \left(1-\frac{M}{2 K M-K R+M}\right)^{1-K},\label{24}\\
\lambda &=&\frac{\lambda _1+\lambda _2-\frac{\lambda _3}{\chi _5(r){}^4 \chi _{35}(r){}^3}}{\lambda _4+\lambda _5-\frac{\lambda _6}{\left(\text{Q1} r^2+1\right)^7 \chi _5(r){}^4 \chi _{35}(r)}}.\label{25}
\end{eqnarray}
where $\lambda _i$, $\{i=1,...,6\}$ are given in the Appendix (\textbf{I}). It is worthwhile to mention here that we have to consider high precision values of parameters as shown in Table-\textbf{I}. The parameters $Q,~
L,~ F,$ and $\lambda$ have been computed for different values of $K$ using Eqs. (\ref{22}-\ref{25}).
In fact, the variations are at very small scale. If we consider, less decimal place accuracy, we can not perform the graphical analysis for different values of parameter $K$.

\section{PHYSICAL PROPERTIES OF THE ANISOTROPIC STELLAR STRUCTURES in $f(\mathscr{R},\mathscr{G})$ Gravity}

In this section , we inquire about the physical properties of stellar structures, Model-1 (LMC X-4) and Model-2 (EXO 1785-248) in $f(\mathscr{R},\mathscr{G})$ gravity with the help of some analytical and graphical analysis. Moreover, in all the figures under discussion, left plot is for Model-1 and Right plot is for Model-2.

\subsection{Evolution of metric functions}

Both metric coefficients $g_{rr}$, and $g_{tt}$ have a crucial role in study of compact stars. Therefore, the behavior of both the metric functions can be noticed from Fig. \textbf{1} for both compact star models in $f(\mathscr{R},\mathscr{G})$ gravity. From Table-\textbf{II}, it is seen that $g_{rr}\mid_{(r =0)}=1$ and $g_{tt}\mid_{(r =0)}\neq0$, which suggests that the Karmarkar condition is physically acceptable to further analyze the configurations of stellar structures. It is also worthwhile to mention here that in current study the $g_{tt}$ component is varied against the different values of parameter $K$, while the $g_{rr}$ component remains fixed as equal to 1.
\begin{figure}
\centering \epsfig{file=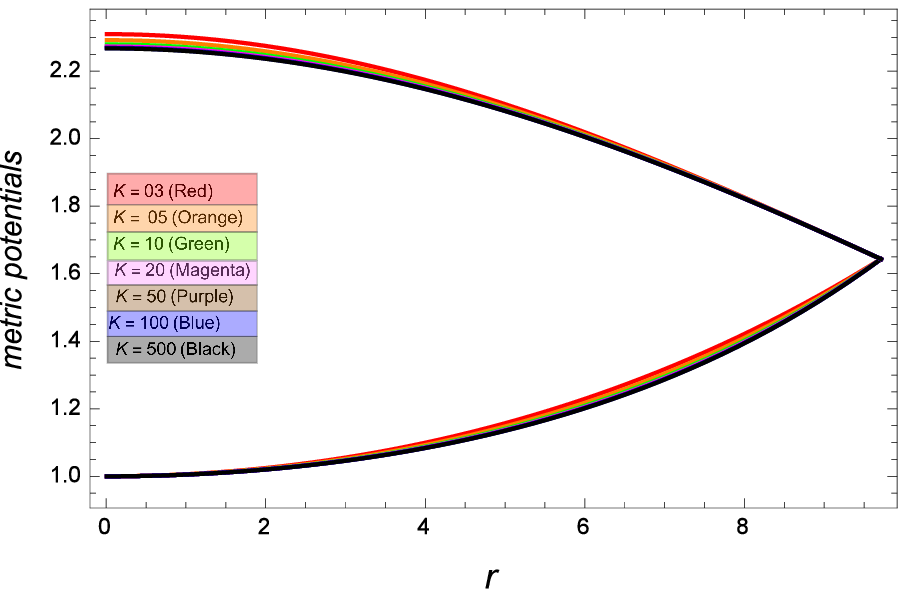, width=.48\linewidth,
height=2.3in}\epsfig{file=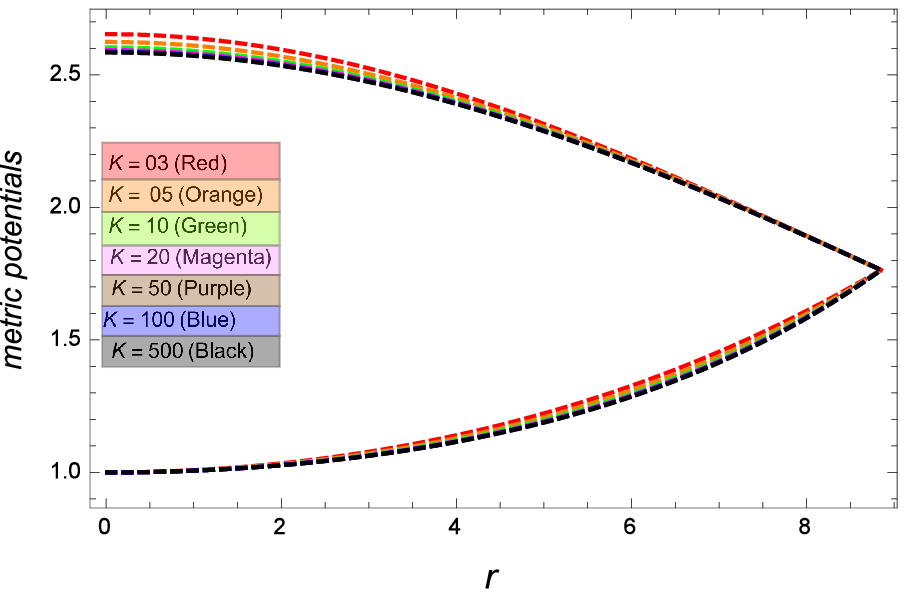, width=.48\linewidth,
height=2.3in}\caption{\label{Fig.1} Evolution of metric functions}
\end{figure}
\begin{figure}
\centering \epsfig{file=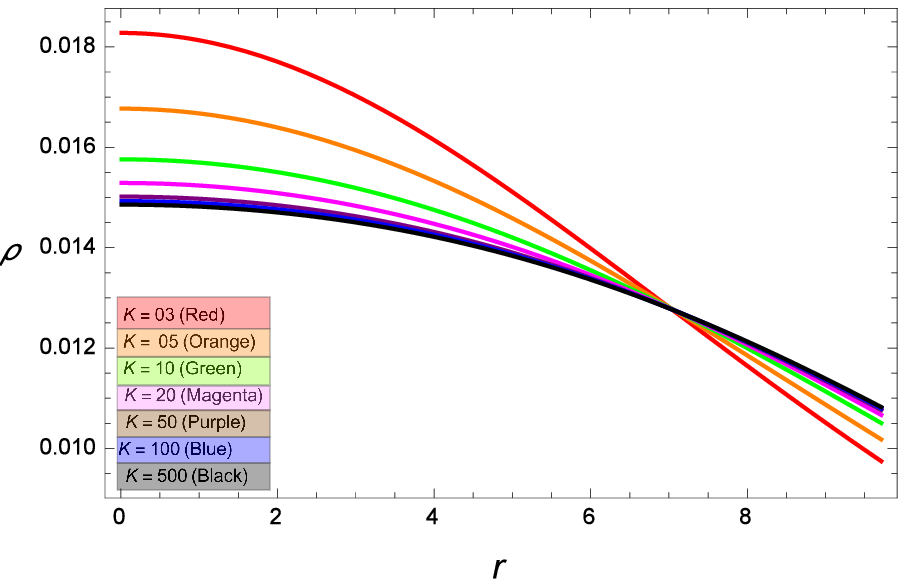, width=.48\linewidth,
height=2.3in}\epsfig{file=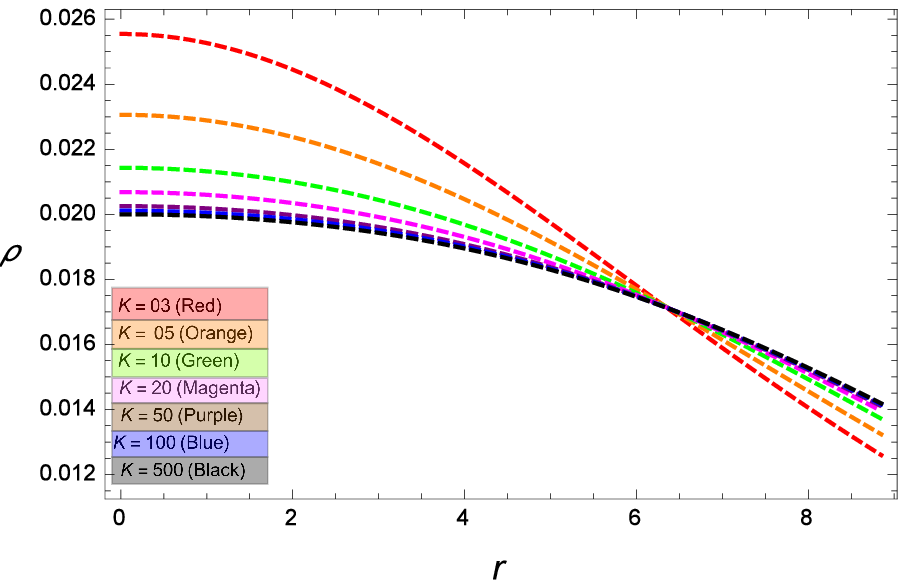, width=.48\linewidth,
height=2.3in}\caption{\label{Fig.2} Behavior of energy density function}
\end{figure}

\subsection{Energy Density and Pressure}

The Fig. \textbf{2} describes the evolution of energy density function for the two compact star models in $f(\mathscr{R},\mathscr{G})$ gravity. It is shown positive throughout the configuration for all values of parameter $K$, for both models. It can be seen from Fig. \textbf{2} and Table-\textbf{II} that the energy density at center is calculated as maximum, then it is varied with decreasing behavior towards the boundary, i.e., $r=R$ and goes to a minimum value. It is argued that energy density function obeys all the physical requirement for the compact stars structures. Moreover, the Zeldovich's condition, i.e. $p_{rc}/ \rho_{c}\leq 1$ is also satisfied as shown in Table-\textbf{II}.

The pressure distribution for anisotropic matter can be divided into two parts, which are known as radial and tangential pressures. Here, we discuss both the pressure components for two models of stellar structures in $f(\mathscr{R},\mathscr{G})$ gravity. It is evident from Fig. \textbf{3} that the radial pressure remains positive for $r<R$. On the other hand the tangential pressure remains positive for both the models throughout the configurations as shown in Fig. \textbf{4}. We calculate both the pressure components, which are seen maximum at center. The radial pressure is vanished at boundary $r=0$, which leads to a positive sign for the stability of both the compact star models. The positivity in tangential pressure at boundary also indicates that both stellar models are stable and physically acceptable in $f(\mathscr{R},\mathscr{G})$ gravity under the Karmarkar condition.
\begin{figure}
\centering \epsfig{file=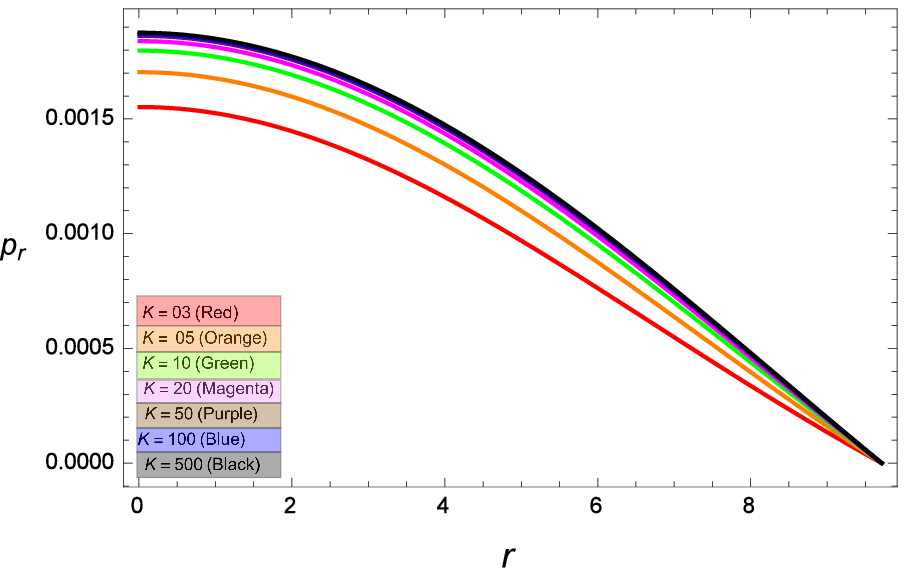, width=.48\linewidth,
height=2.3in}\epsfig{file=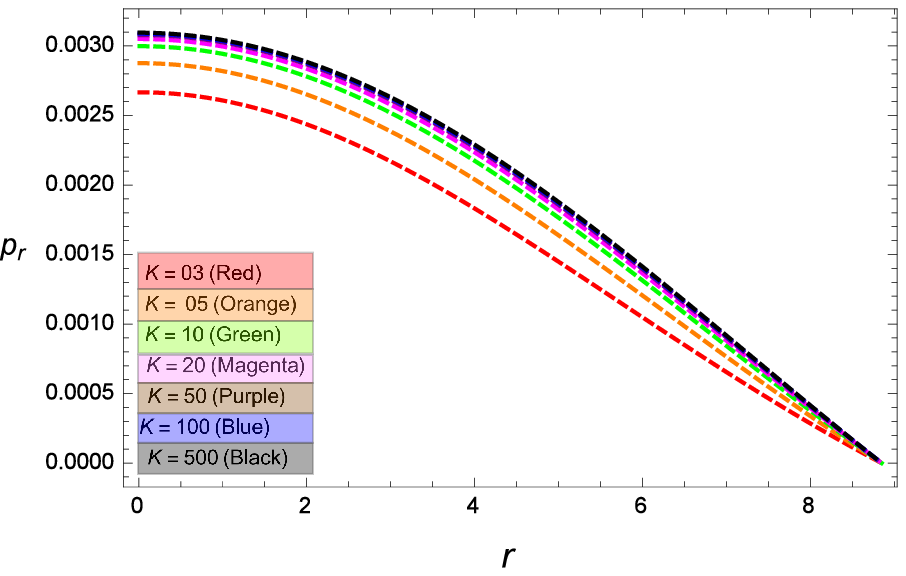, width=.48\linewidth,
height=2.3in}
\caption{\label{Fig.3} Behavior of radial pressure}
\end{figure}
\begin{figure}
\centering \epsfig{file=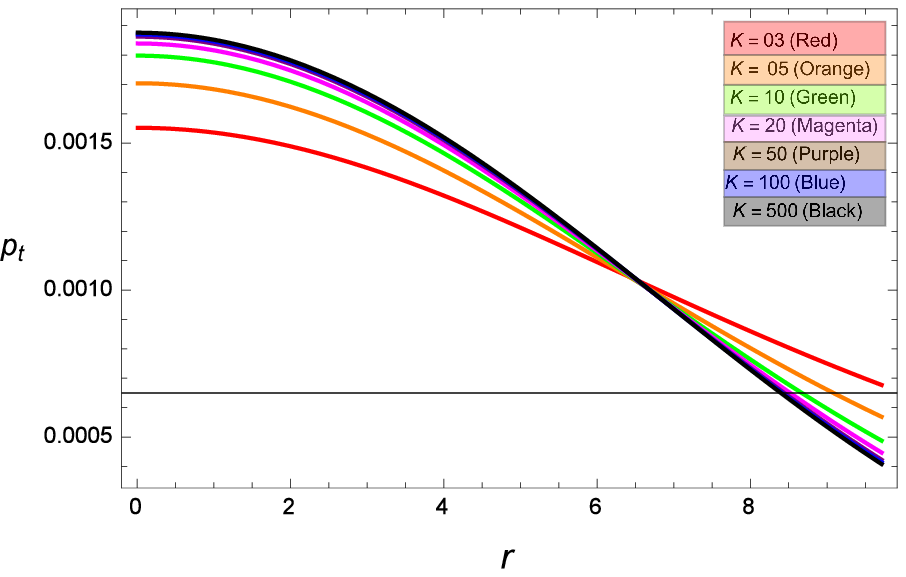, width=.48\linewidth,
height=2.3in}\epsfig{file=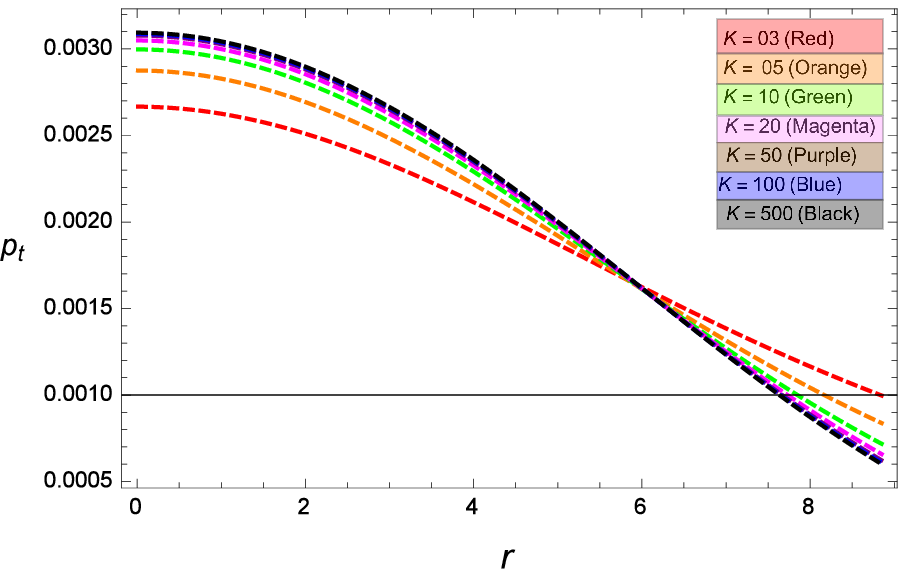, width=.48\linewidth,
height=2.3in}\caption{\label{Fig.4} Behavior of tangential pressure}
\end{figure}

\subsection{Anisotropy and Gradients}

For massive stellar objects, the radial pressure may not be equal to the tangential one. Different arguments have been given for the existence of anisotropy in stellar models such as by the presence of type $3$A superfluid \cite{8143} and different kinds of phase
transitions \cite{9143}. In fact, anisotropy is also important to understand the peculiar properties of matter in the core of stellar structure.
The difference of pressure components leads to the concept of anisotropy function, i.e., $\triangle=p_{t}-p_{r}$. It is observed from Table-\textbf{II} that both the pressure components have same values at center, i.e., $r=0$, but tangential pressure remains positive at boundary while the radial pressure vanishes at boundary, which indicates that the anisotropy function remains positive throughout the configuration. In current study, the anisotropy function is shown positive with regularly increasing behavior for both the models as shown in Fig. \textbf{5}. It is evident that the anisotropy function is zero at center due to equal values of radial and tangential values at center then it monotonically increases towards boundary and becomes maximum at $r=R$. This trend in the values of anisotropy function shows that our calculated results satisfy the stability conditions for both the stellar configurations.
\begin{figure}
\centering \epsfig{file=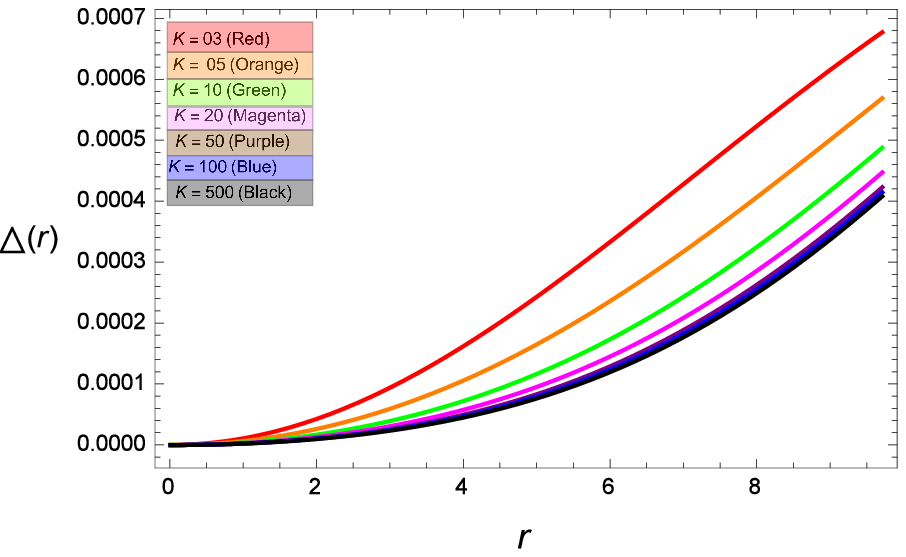, width=.48\linewidth,
height=2.3in}\epsfig{file=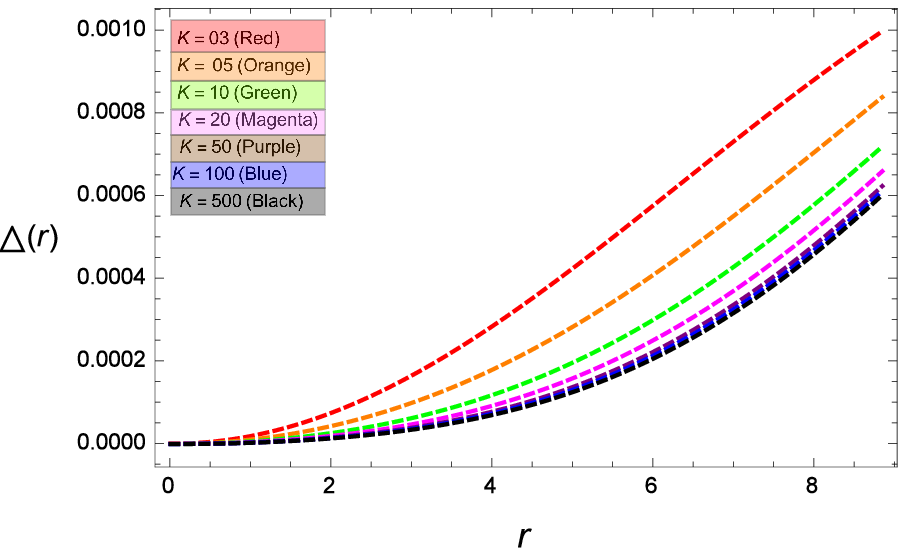, width=.48\linewidth,
height=2.3in}\caption{\label{Fig.5} Behavior of anisotropy function}
\end{figure}
\begin{figure}
\centering  \epsfig{file=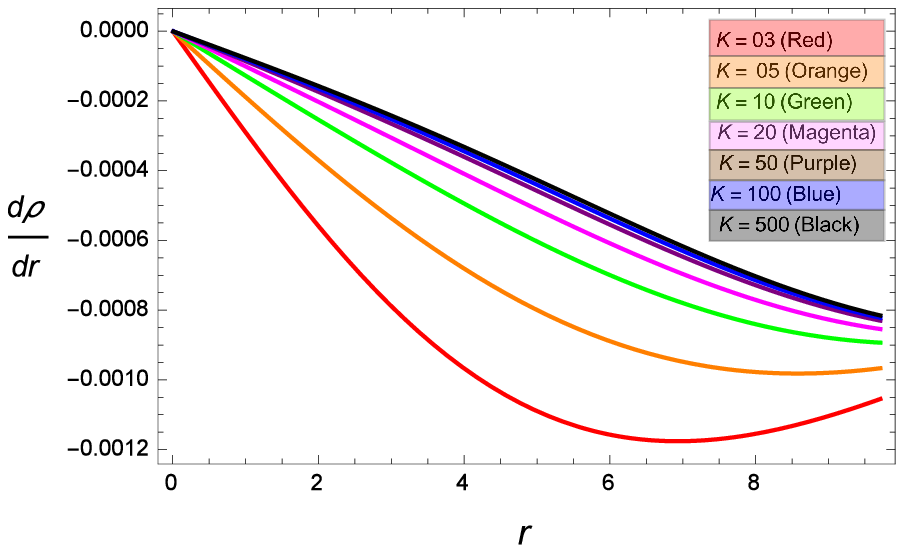, width=.48\linewidth,
height=2.3in}\epsfig{file=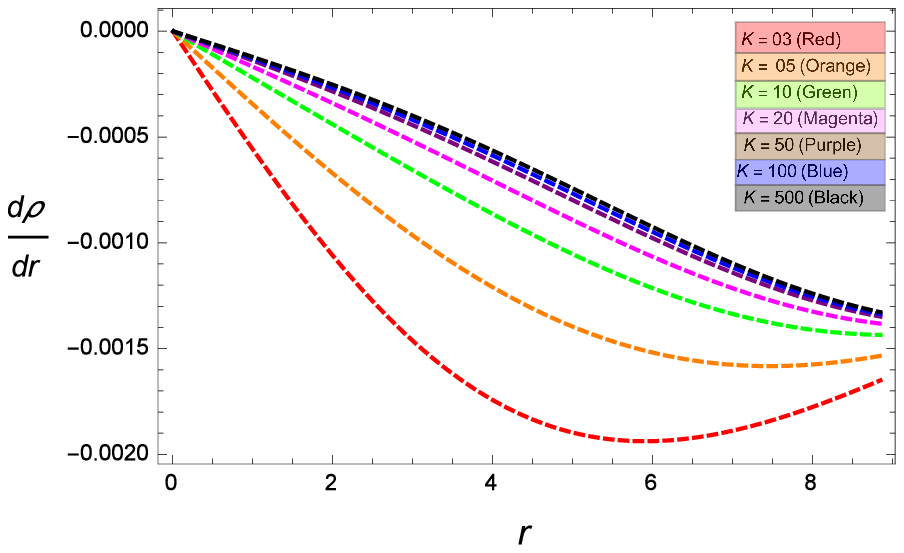, width=.48\linewidth,
height=2.3in}\caption{\label{Fig.6} Behavior of derivative of energy density}
\end{figure}

Now we analyze the calculus for the current study, i.e., first order derivatives $d\rho/dr, \;dp_{r}/dr$ and $d p_{t}/dr$. The following relations must be satisfied
\begin{equation*}
\frac{d\rho}{dr}< 0,\;\;\;\;\;\;\; \;\frac{dp_{r}}{dr}< 0,\;\;\;\;\;\;\;\frac{dp_{t}}{dr}< 0.
\end{equation*}
The above inequalities can be seen verified from the Figs. (\textbf{6}-\textbf{8}). The negative nature of these gradients is assumed as a necessary condition for the stellar modeling. Further, these gradients vanish at $r=0$, i.e.,
\begin{equation*}
\frac{d\rho}{dr}\mid_{r=0}= 0,\;\;\;\;\;\;\; \;\frac{dp_{r}}{dr}\mid_{r=0}= 0,\;\;\;\;\;\;\;\frac{dp_{t}}{dr}\mid_{r=0}= 0.
\end{equation*}
\begin{figure}
\centering \epsfig{file=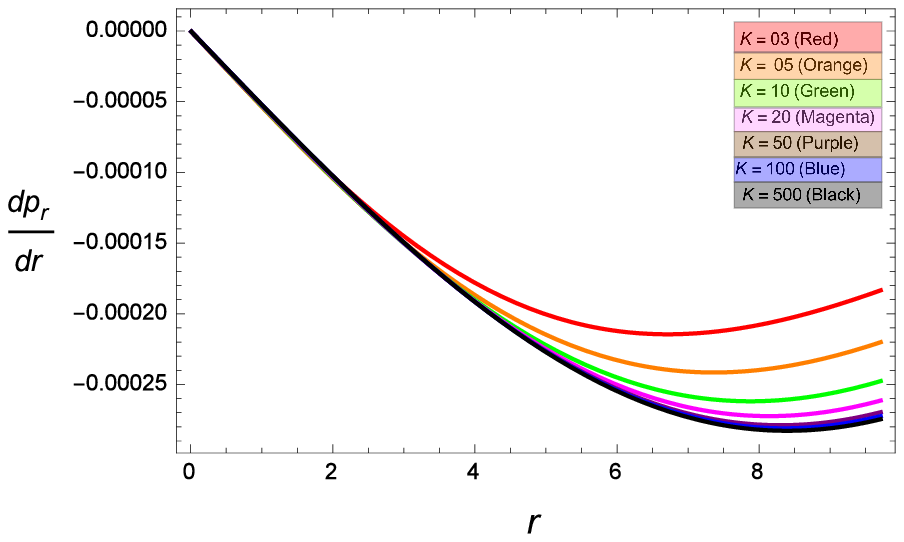, width=.48\linewidth,
height=2.3in}\epsfig{file=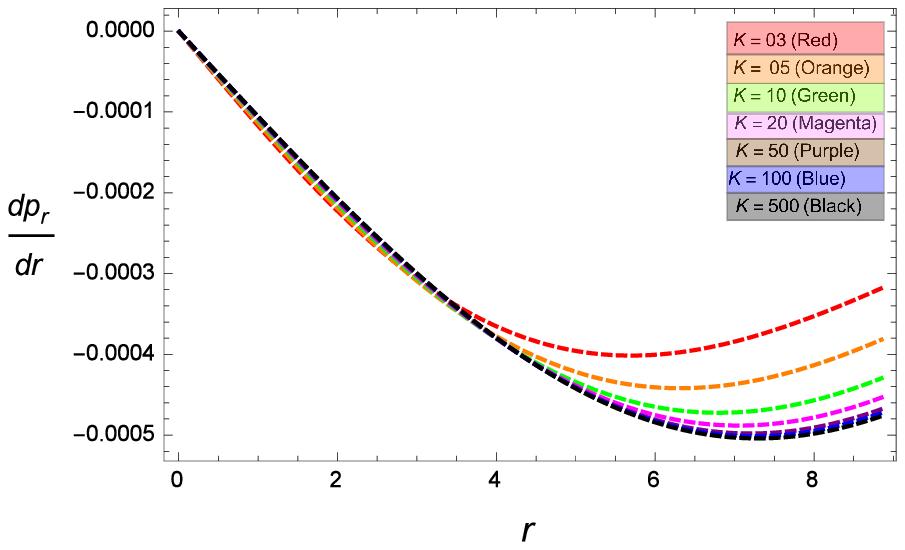, width=.48\linewidth,
height=2.3in} \caption{\label{Fig.7} Behavior of of derivative of radial pressure}
\end{figure}
\begin{figure}
\centering \epsfig{file=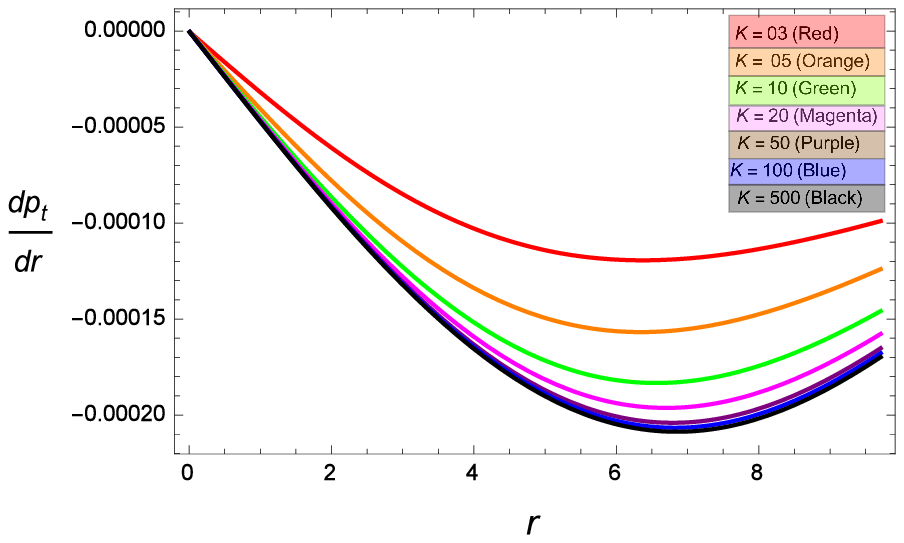, width=.48\linewidth,
height=2.3in}\epsfig{file=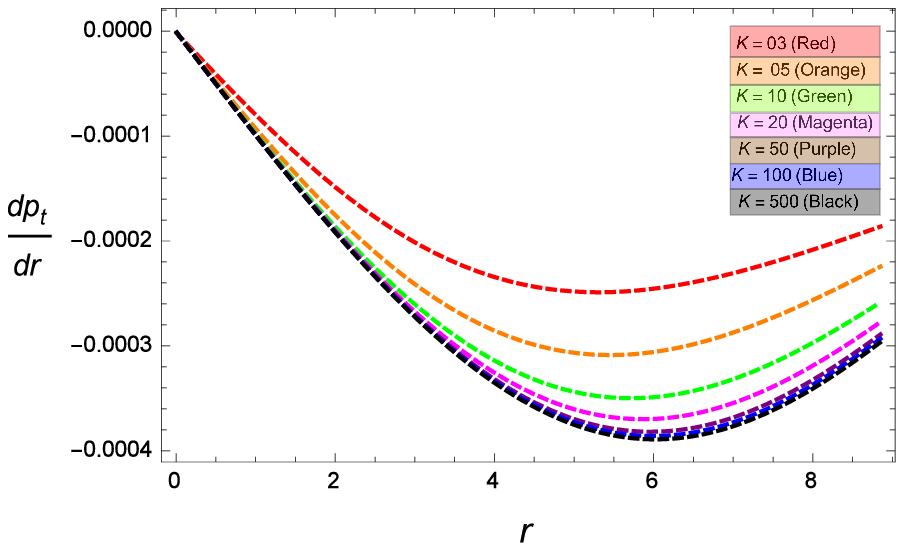, width=.48\linewidth,
height=2.3in}\caption{\label{Fig.8} Behavior of derivative of tangential pressure}
\end{figure}

\subsection{Energy Conditions and Equation of State}

Energy conditions have a significant role in relativistic cosmology. Commonly there are four types of significant energy conditions, which are important in the context of modified theories. These conditions are mentioned as dominant energy condition ($DEC$), strong energy condition ($SEC$), weak energy condition ($WEC$) and null energy condition ($NEC$), and defined as
\begin{eqnarray}\nonumber
NEC: \quad&& \rho\geq0,\;\;\;\;\;\;\;\;\; \;\;\;\;\;\;\;\;\;\;\;\;\;\;\;WEC: \quad\rho-p_{t}\geq0, \quad \rho-p_{r}\geq0,\nonumber\\
SEC: \quad&& \rho-p_{r}-2p_{t}\geq0,\;\;\;\;\;\;\;DEC: \quad\rho>|p_r|, \quad\rho>|p_t|.\nonumber
\end{eqnarray}
The validity of $NEC$ can be confirmed from Fig. \textbf{2}. $WEC$ can be seen obeyed from Figs. \textbf{9} and \textbf{10}. $SEC$ is also satisfied in current study as depicted from Fig. \textbf{11}. As far as $DEC$ is concerned, it can be seen validated from Figs. (\textbf{2}-\textbf{4}).

Equation of state parameters $w_r$ and $w_t$, can be calculated by the following relations
\begin{equation}\label{26}
w_r\times \rho =p_r,\;\;\;\;\;\;\;\;\;\;\;\;\;\;\;\;\;\;\;\;w_t\times \rho =p_t.
\end{equation}
The evolution of $w_r$ and $w_t$ can be seen from Fig. \textbf{12}. Both the important ratios lie in the interval $(0,1)$. Thus our obtained solutions are physically acceptable in the background of Karmarkar condition in $f(\mathscr{R},\mathscr{G})$ gravity.\newpage
\begin{center}
\begin{table}
\caption{\label{tab1}{Calculated values of different physical parameters at center and boundary.}}
\begin{tabular}{|c|c|c|c|c|c|c|c|c|}
    \hline
    & \multicolumn{7}{|c|}{LMC X-4} \\
\hline
K     &$e^{a(r=0)}$   &$e^{b(r=0)}$   &$\rho_{R}\;(g/cm^3)$      &$p_{r_{c}}\;(dyne/cm^{2})$     &$p_{t_{c}}\;(dyne/cm^{2})$   &$\rho_{c}\;(g/cm^3)$           &$p_{r_{c}}/\rho_{c}=p_{t_{c}}/\rho_{c}$\\
\hline
3     &0.43294        &1.0            &$0.97462\times 10^{15}$    &$15.5221\times 10^{35}$        &$15.5221\times 10^{35}$      &$1.82789\times 10^{15}$        &$0.084918$\\
\hline
5     &0.43632        &1.0            &$1.01798\times 10^{15}$    &$17.0387\times 10^{35}$        &$17.0387\times 10^{35}$      &$1.67679\times 10^{15}$        &$0.101622$\\
\hline
10     &0.43874        &1.0            &$1.05050\times 10^{15}$    &$17.9789\times 10^{35}$       &$17.9789\times 10^{35}$      &$1.57629\times 10^{15}$        &$0.114058$\\
\hline
20     &0.43992        &1.0            &$1.06676\times 10^{15}$    &$18.3935\times 10^{35}$       &$18.3935\times 10^{35}$      &$1.52926\times 10^{15}$        &$0.120263$\\
\hline
50     &0.44061       &1.0            &$1.07652\times 10^{15}$    &$18.6264\times 10^{35}$       &$18.6264\times 10^{35}$      &$1.50178\times 10^{15}$        &$0.124029$\\
\hline
100     &0.44084        &1.0            &$1.07977\times 10^{15}$    &$18.7016\times 10^{35}$       &$18.7016\times 10^{35}$      &$1.49283\times 10^{15}$        &$0.125258$\\
\hline
500     &0.44102       &1.0            &$1.08238\times 10^{15}$    &$18.7580\times 10^{35}$       &$18.7580\times 10^{35}$      &$1.48583\times 10^{15}$        &$0.126266$\\
\hline
 & \multicolumn{7}{|c|}{EXO 1785-248} \\
\hline
3     &0.37688        &1.0            &$1.25987\times 10^{15}$    &$26.6655\times 10^{35}$        &$26.6655\times 10^{35}$      &$2.55345\times 10^{15}$        &$0.104429$\\
\hline
5     &0.38113        &1.0            &$1.32375\times 10^{15}$    &$28.7635\times 10^{35}$        &$28.7635\times 10^{35}$      &$2.30535\times 10^{15}$        &$0.124736$\\
\hline
10     &0.38415        &1.0            &$1.37166\times 10^{15}$    &$29.9850\times 10^{35}$       &$29.9850\times 10^{35}$      &$2.14417\times 10^{15}$        &$0.139803$\\
\hline
20     &0.38561        &1.0            &$1.39562\times 10^{15}$    &$30.5032\times 10^{35}$       &$30.5032\times 10^{35}$      &$2.06938\times 10^{15}$        &$0.147403$\\
\hline
50     &0.38647       &1.0            &$1.41001\times 10^{15}$    &$30.7891\times 10^{35}$       &$30.7891\times 10^{35}$      &$2.02555\times 10^{15}$        &$0.151947$\\
\hline
100     &0.38675        &1.0            &$1.41481\times 10^{15}$    &$30.8813\times 10^{35}$       &$30.8813\times 10^{35}$      &$2.01150\times 10^{15}$        &$0.153524$\\
\hline
500     &0.38698       &1.0            &$1.41865\times 10^{15}$    &$30.9404\times 10^{35}$       &$30.9404\times 10^{35}$      &$1.99894\times 10^{15}$        &$0.154855$\\
\hline
\end{tabular}
\end{table}
\end{center}
\begin{figure}
\centering \epsfig{file=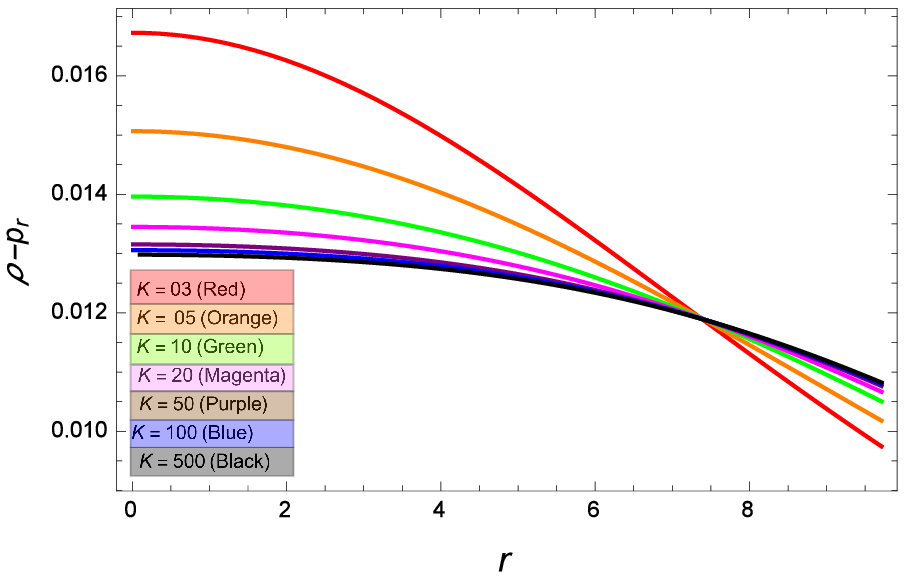, width=.48\linewidth,
height=2.3in}\epsfig{file=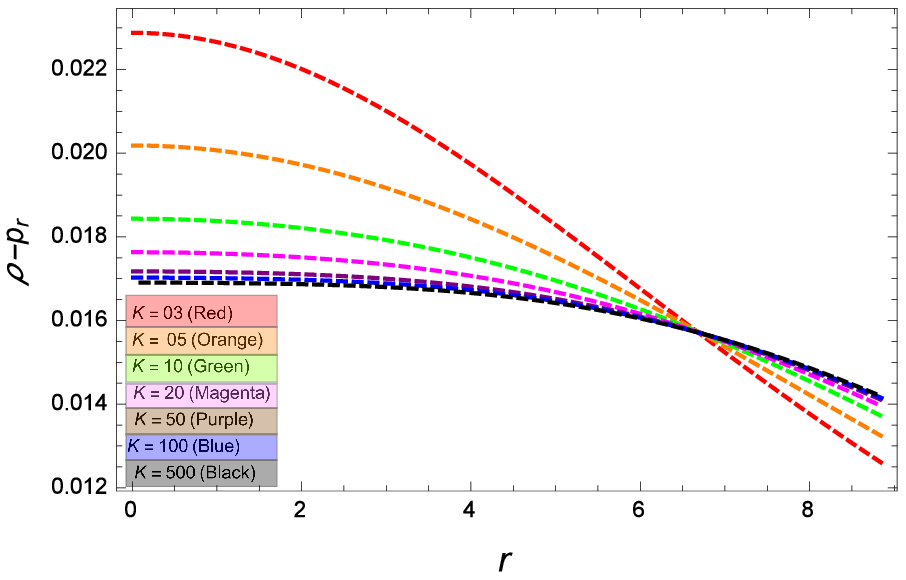, width=.48\linewidth,
height=2.3in} \caption{\label{Fig.9} Behavior of $\rho-p_{r}$}
\end{figure}
\begin{figure}
\centering \epsfig{file=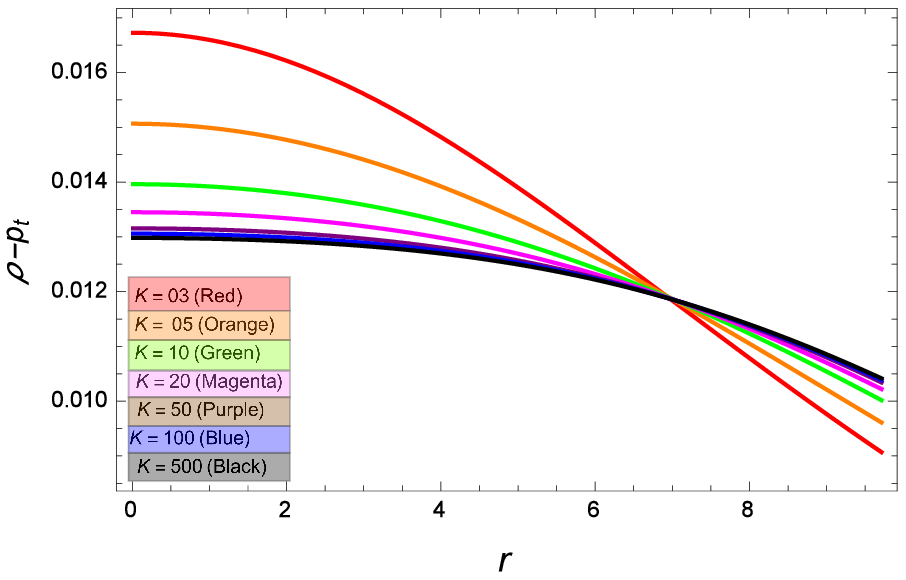, width=.48\linewidth,
height=2.3in}\epsfig{file=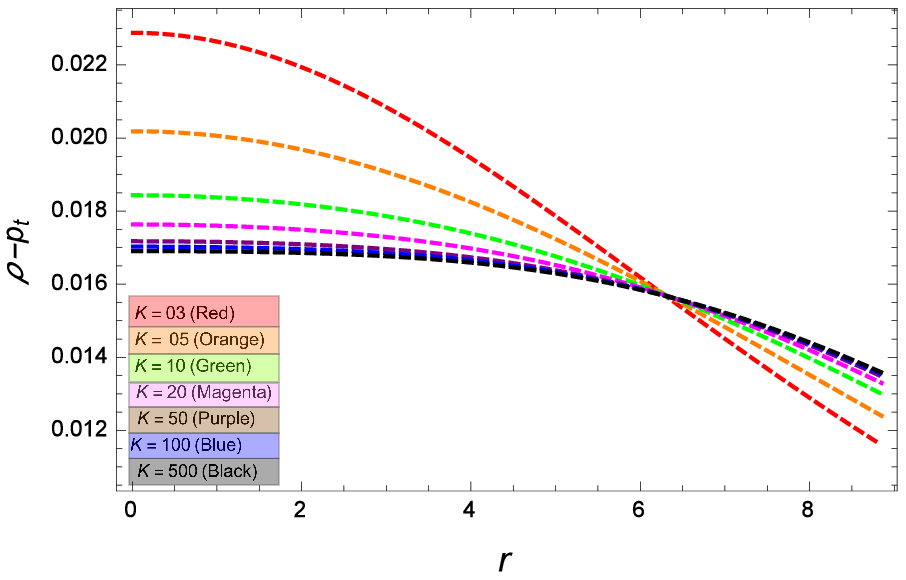, width=.48\linewidth,
height=2.3in}\caption{\label{Fig.10} Behavior of $\rho-p_{t}$}
\end{figure}
\begin{figure}
\centering \epsfig{file=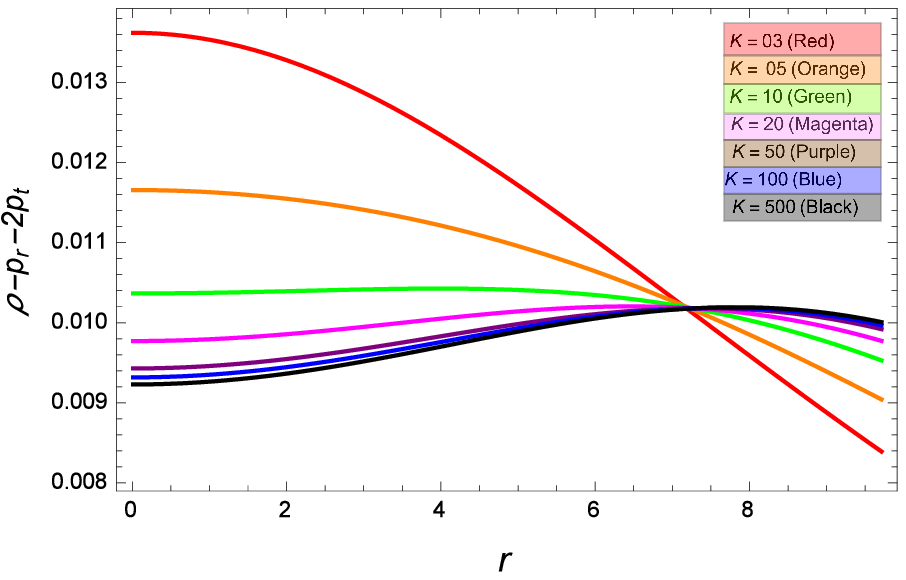, width=.48\linewidth,
height=2.3in}\epsfig{file=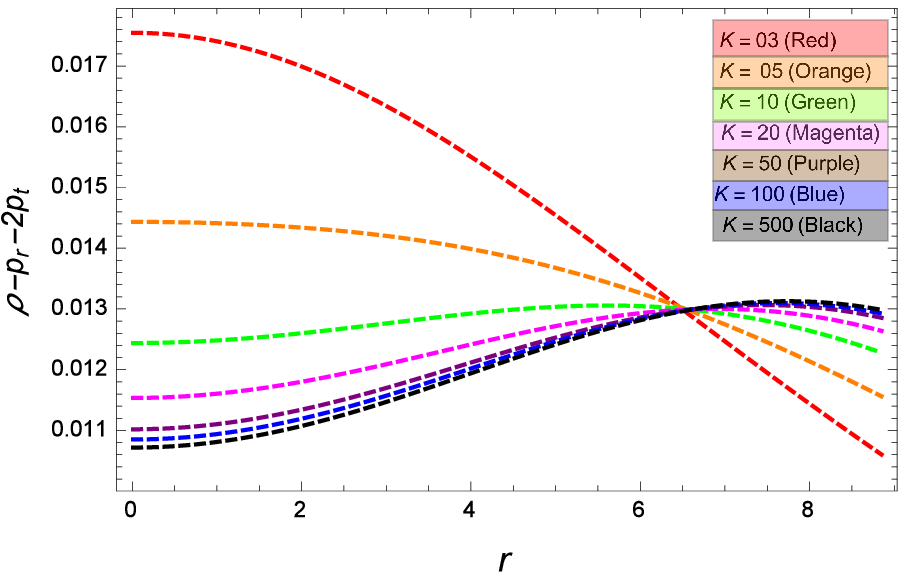, width=.48\linewidth,
height=2.3in} \caption{\label{Fig.11} Behavior of $\rho-p_{r}-2p_{t}$}
\end{figure}
\begin{figure}
\centering \epsfig{file=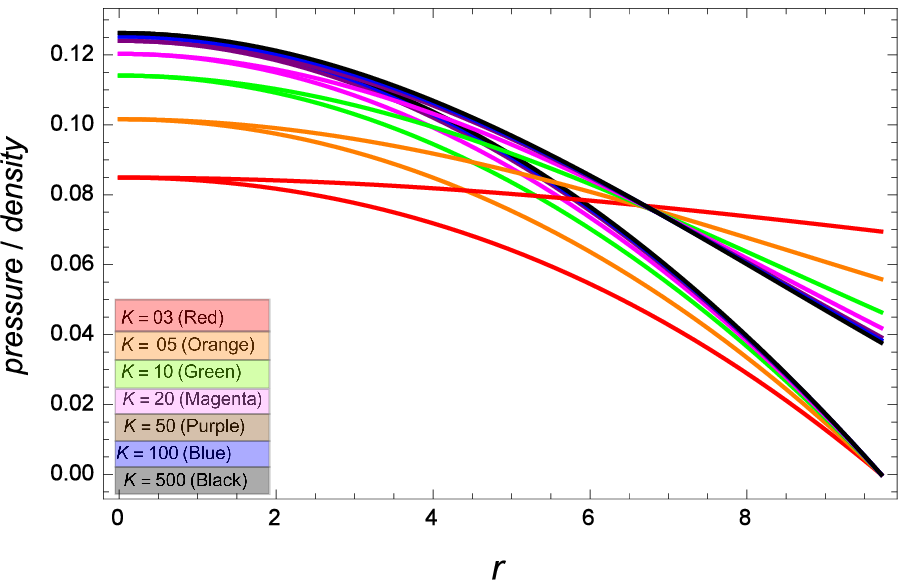, width=.48\linewidth,
height=2.3in}\epsfig{file=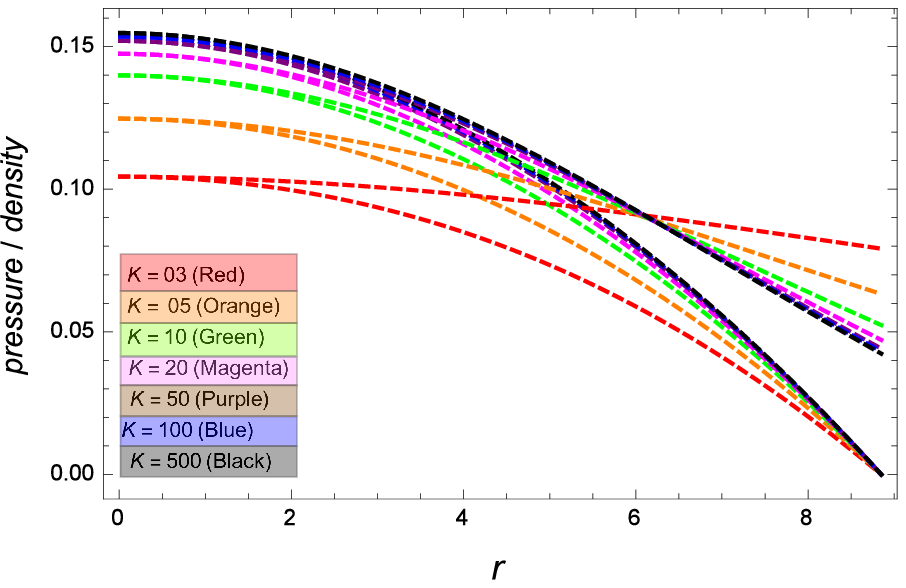, width=.48\linewidth,
height=2.3in}\caption{\label{Fig.12} Behavior of $w_r$ and $w_t$}
\end{figure}

\subsection{Gravitational red-shift function, compactness function, and mass function}

The gravitational red-shift function for stellar objects is presented as
\begin{equation}\label{27}
Z_{s}=\frac{1}{\sqrt{1-2u(r)}}\left(1-\sqrt{1-2u(r)}\right),
\end{equation}
where $u(r)$ mentions the compactness function of compact stars, defined by the following expression
\begin{equation}\label{28}
u(r)=\frac{2}{r}\times m(r).
\end{equation}
Here $m(r)$ denotes the mass-function of stellar objects defined as
\begin{equation}\label{29}
m(r)=4\pi \times\int^{r}_{0}(r^{2}\times \rho)dr
\end{equation}
\begin{figure}
\centering \epsfig{file=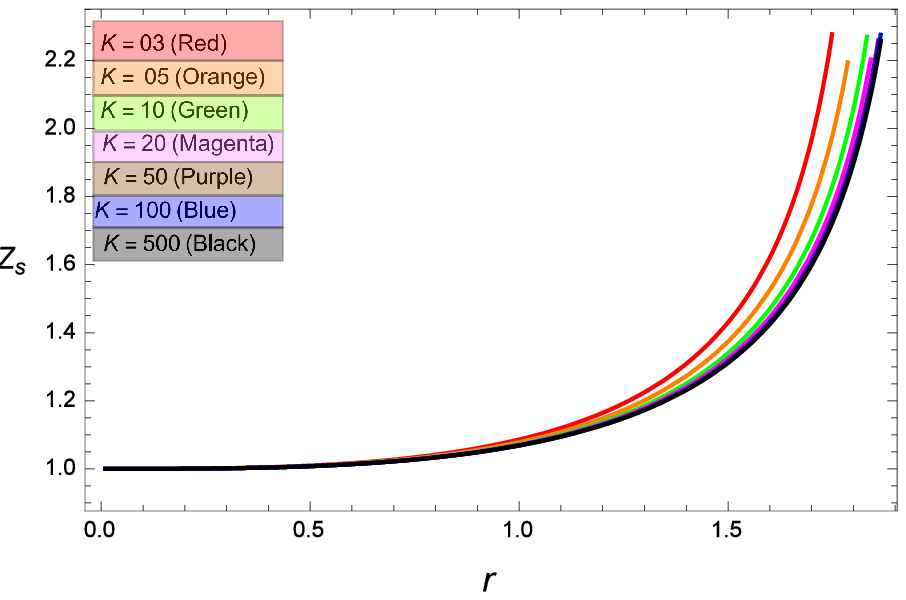, width=.48\linewidth,
height=2.3in}\epsfig{file=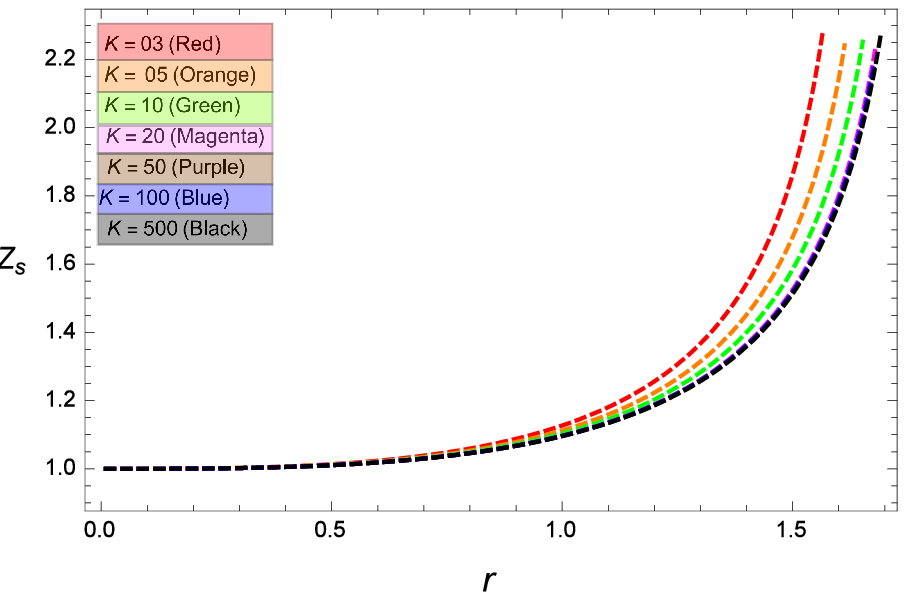, width=.48\linewidth,
height=2.3in} \caption{\label{Fig.13} Evolution of $Z_{s}$}
\end{figure}
\begin{figure}
\centering \epsfig{file=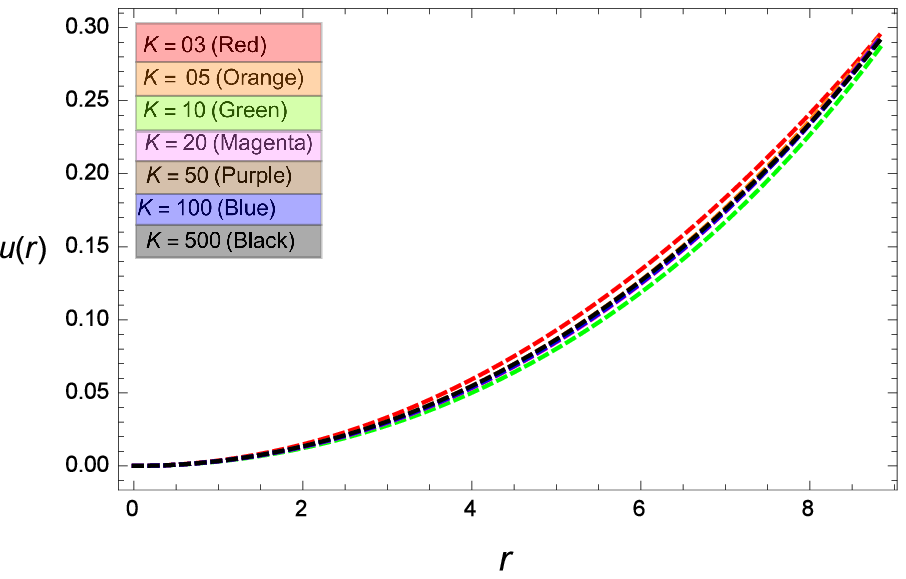, width=.48\linewidth,
height=2.3in}\epsfig{file=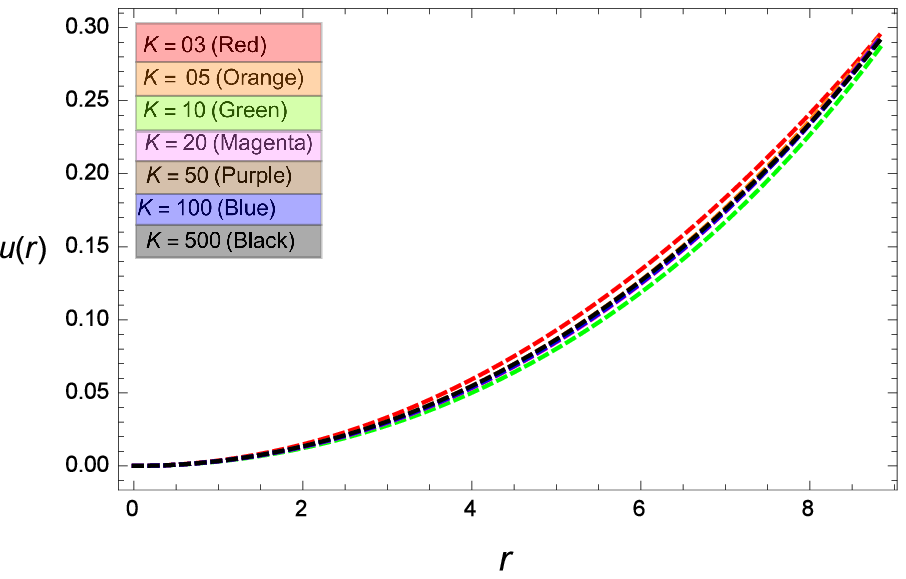, width=.48\linewidth,
height=2.3in}\caption{\label{Fig.14} Evolution of $u(r)$.}
\end{figure}
\begin{figure}
\centering \epsfig{file=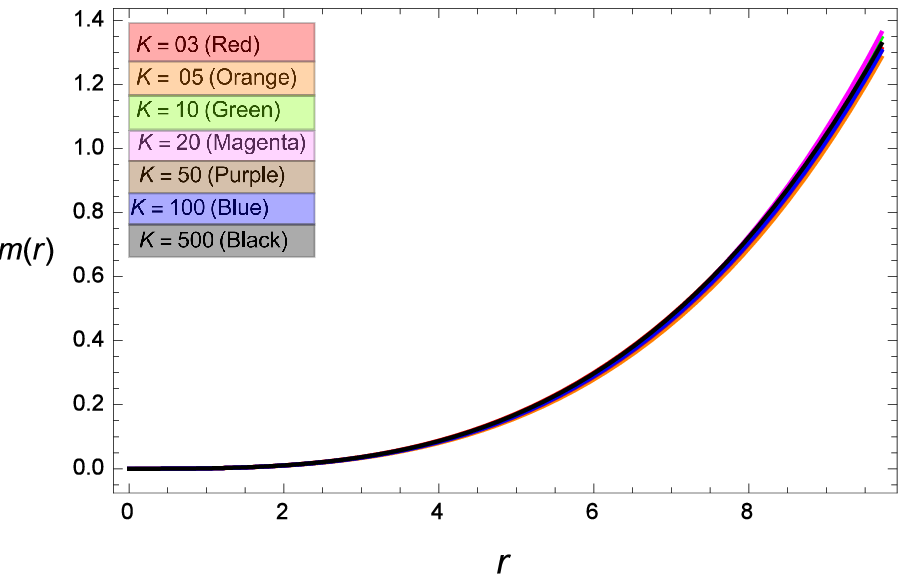, width=.48\linewidth,
height=2.3in}\epsfig{file=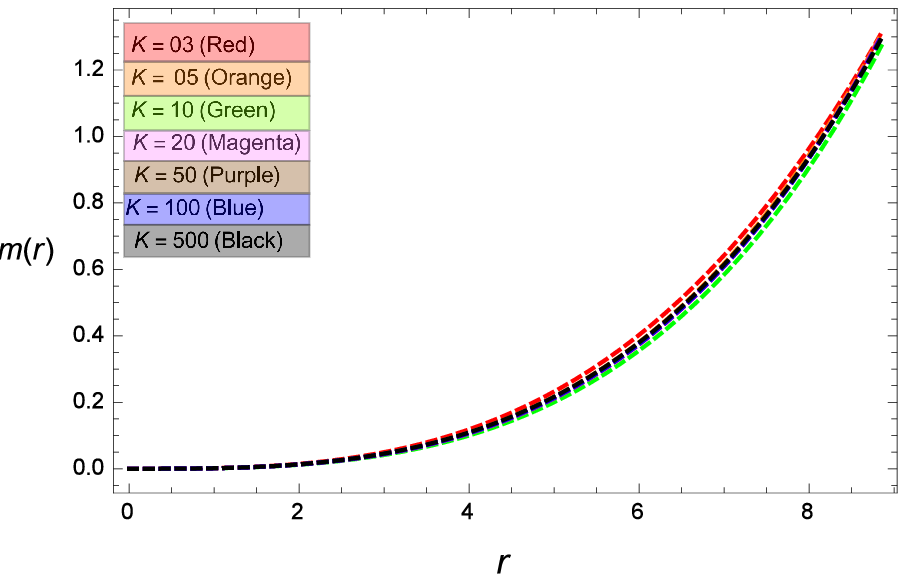, width=.48\linewidth,
height=2.3in} \caption{\label{Fig.15} Evolution of $m(r)$.}
\end{figure}
The above three physical parameters play an important role for stellar modeling. The behavior of red-shift parameter $Z_{s}$, can be perceived from Fig. \textbf{13} for both the models. For an anisotropic fluid sphere, Bohmer and Harko \cite{44} proposed that $Z_{s}$ should remain less than five, i.e., $Z_{s}\leq 5$ and Ivanov \cite{45} suggested that  $Z_{s}\leq 5.211$. Our current study reveals that $Z_{s}\leq 2.400$, showing the validity of stellar structures in modified $f(\mathscr{R},\mathscr{G})$ gravity. As far as the compactness function $u(r)$ is concerned, it can be seen from Fig. \textbf{14}, that $u(r)\leq 0.30$, which implies that Buchdahl condition \cite{46} is also satisfied for the current analysis. The mass-function $m(r)$, is plotted in Fig. \textbf{15} and it is evident that the mass-function is monotonically increasing toward the surface.

\subsection{Equilibrium Condition}

Here we discuss the equilibrium configuration of the stellar structure models in $f(\mathscr{R},\mathscr{G})$ gravity. For this purpose, we investigate the TOV equation. For the given spherically symmetric space-time, TOV equation is given by
\begin{equation}\label{30}
\frac{2}{r}(p_{t}-p_{r})-\frac{d p _{r}}{dr}-\frac{a(r)'}{2}(\rho+p_{r})=0.
\end{equation}
The above equation can be symbolized as
\begin{equation}\label{31}
\mathscr{F}_{\mathrm{h}}+\mathscr{F}_{\mathrm{g}}+\mathscr{F}_{\mathrm{a}}=0,
\end{equation}
such that
\begin{equation*}
\mathscr{F}_{\mathrm{a}}=\frac{2}{r}(p_{t}-p_{r}),\;\;\;\;\;\;\mathscr{F}_{\mathrm{h}}=-\frac{d p _{r}}{dr},\;\;\;\;\;\;\; \mathscr{F}_{\mathrm{g}}=-\frac{a(r)'}{2}(\rho+p_{r}),
\end{equation*}
where $\mathscr{F}_{\mathrm{a}}$, $\mathscr{F}_{\mathrm{h}}$, and $\mathscr{F}_{\mathrm{g}}$ denote the anisotropic force, hydrostatic force, and gravitational force respectively. The balancing feature of anisotropic, hydrostatic, and gravitational forces can be seen from Fig. \textbf{16}. This balancing nature of $\mathscr{F}_{\mathrm{a}}$, $\mathscr{F}_{\mathrm{h}}$, and $\mathscr{F}_{\mathrm{g}}$ demonstrate that stellar structures under discussion are seen stable and physically acceptable.
\begin{figure}
\centering \epsfig{file=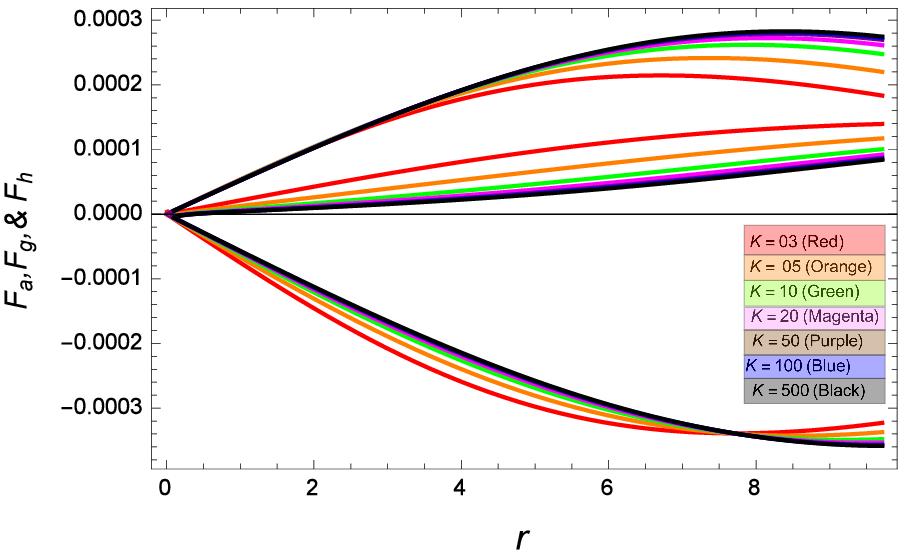, width=.48\linewidth,
height=2.3in}\epsfig{file=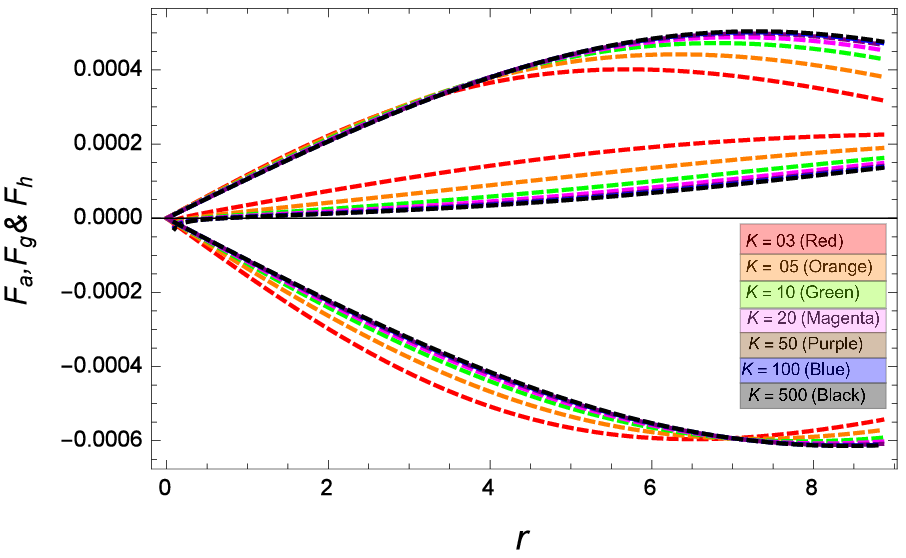, width=.48\linewidth,
height=2.3in}\caption{\label{Fig.16} Balancing behavior of $\mathscr{F}_{\mathrm{a}}$, $\mathscr{F}_{\mathrm{h}}$, and $\mathscr{F}_{\mathrm{g}}$}
\end{figure}

\subsection{Stability Analysis: Causality Condition and Adiabatic index}

In order to discuss the well-known stability criteria, i.e., causality condition, we investigate two kinds of speeds of sound, i.e., radial and tangential speeds of sound, which are mentioned by $\nu^{2}_{r}$ and $\nu^{2}_{t}$ and calculated as
\begin{equation}\label{32}
\nu_{r}= \sqrt{\frac{dp_{r}}{d\rho}}\;\;\;\;\;\;\Rightarrow  \nu^{2}_{r}= \frac{dp_{r}}{dr} \times \frac{dr}{d\rho},\;\;\;\;\;\;\;\;\;\;\nu_{t}= \sqrt{\frac{dp_{t}}{d\rho}} \;\;\;\;\;\;\Rightarrow  \nu^{2}_{t}= \frac{dp_{t}}{dr} \times \frac{dr}{d\rho}.
\end{equation}
\begin{figure}
\centering \epsfig{file=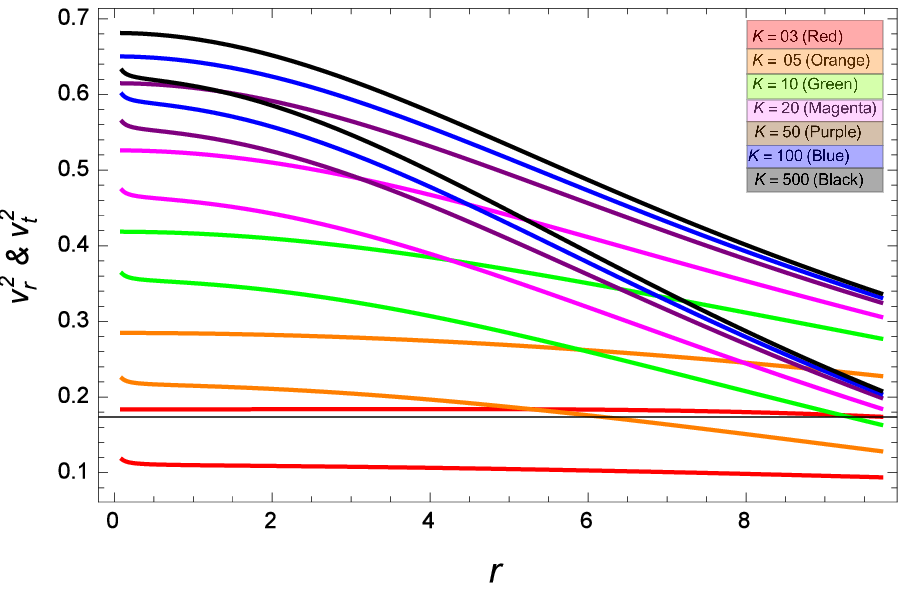, width=.48\linewidth,
height=2.3in}\epsfig{file=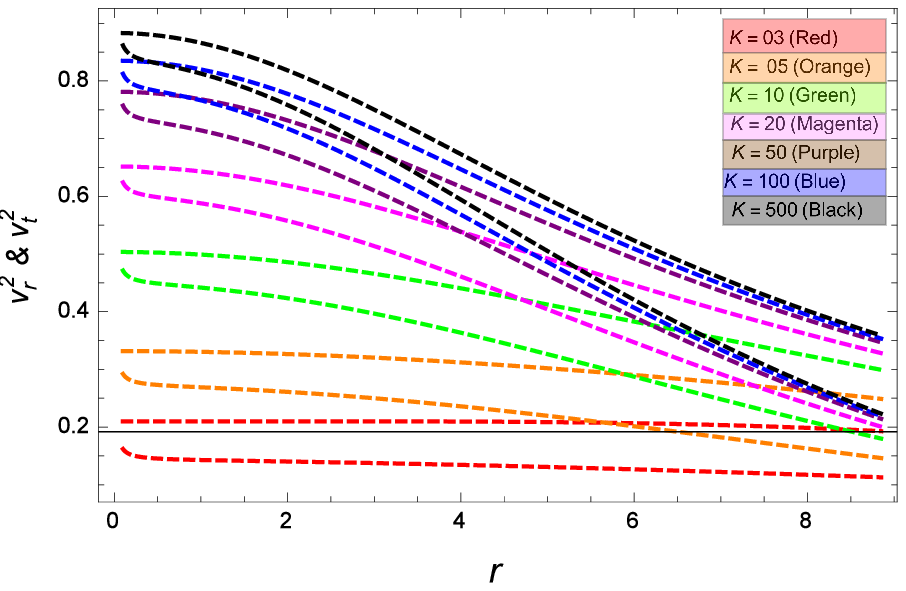, width=.48\linewidth,
height=2.3in} \caption{\label{Fig.17} Behavior of $\nu^{2}_{r}$ and $\nu^{2}_{t}$.}
\end{figure}
\begin{figure}
\centering \epsfig{file=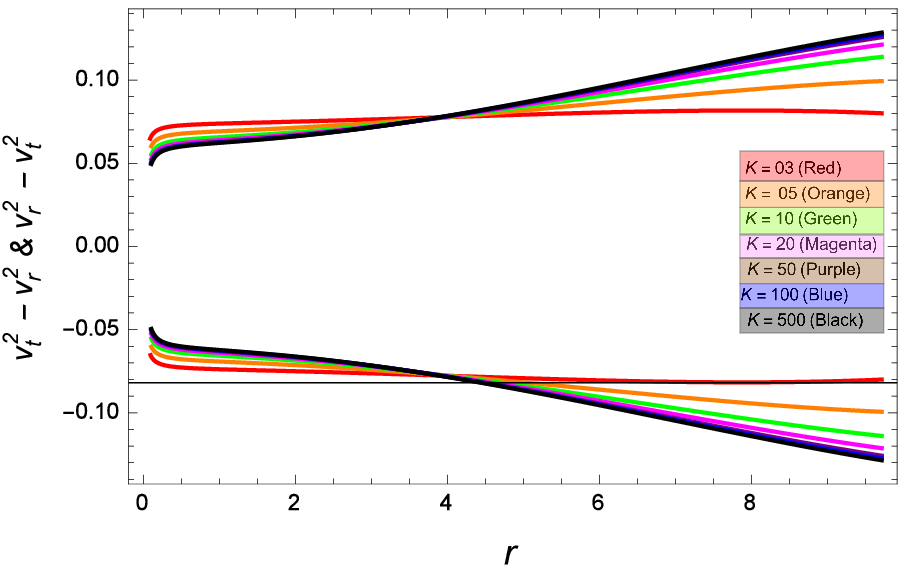, width=.48\linewidth,
height=2.3in}\epsfig{file=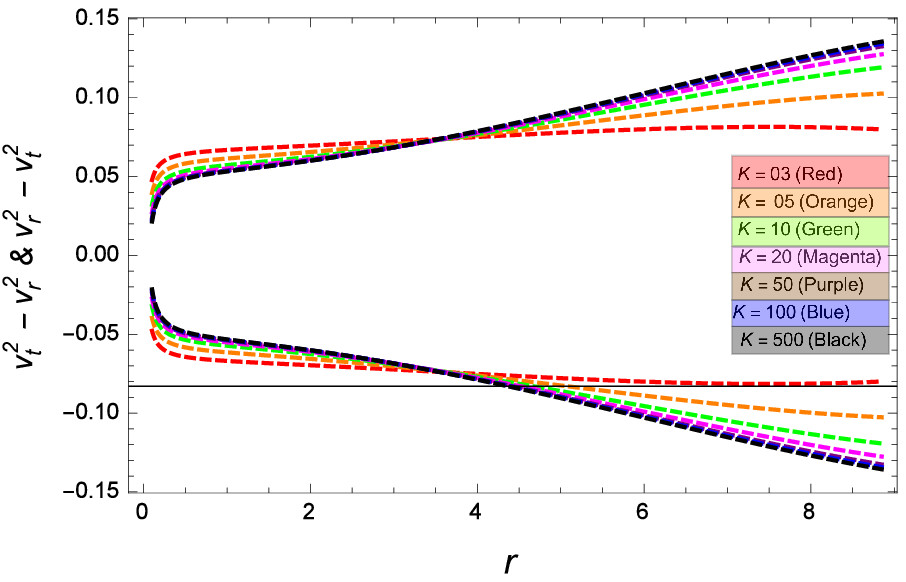, width=.48\linewidth,
height=2.3in}\caption{\label{Fig.18} Abrea condition, i.e., $-1\leq \nu^{2}_{t}-\nu^{2}_{r}\leq 0$ and $0\leq\nu^{2}_{r}-\nu^{2}_{t}\leq 1$}
\end{figure}
It is seen in Fig. \textbf{17}, that both the radial and tangential velocities satisfy the condition $0\leq \nu_{r}~ \& ~\nu_{t}\leq1$. It is also evident in Fig. \textbf{18}, that the Abrea condition \cite{47}, i.e., $-1\leq \nu^{2}_{t}-\nu^{2}_{r}\leq 0$ is also fulfilled. Validity of both conditions confirms that our presented compact star models are potentially stable. The converse Abrea condition, i.e., $0\leq\nu^{2}_{r}-\nu^{2}_{t}\leq 1$, is also seen to be satisfied as shown in Fig. \textbf{18}.

Hillebrandt and Steinmetz \cite{48} presented an important parameter in the context of stellar models with anisotropic fluid
\begin{equation}\label{33}
\Gamma_{r}= \frac{\rho\times \nu^{2}_{r}}{p_{r}}\times \left(1+\frac{p_{r}}{\rho}\right).
\end{equation}
The parameter $\Gamma_{r}$ reveals the stability of Newtonian anisotropic sphere with $\Gamma_{r}>4/3$. An un-stable anisotropic sphere is assumed when $\Gamma_{r}<4/3$. Another way of describing the stability condition for an anisotropic compact structure is given by
\begin{equation}\label{34}
\Gamma_{r}> \frac{4}{3}+\frac{1}{|p^{'}_{r_{c}}|}\left(\frac{\varrho\times \rho_{c}\times p_{r_{c}}}{2}\times r+\frac{4\times(p_{t_{c}}-p_{r_{c}})}{3r}\right),
\end{equation}
where $\varrho$ is a real number.
\begin{figure}
\centering \epsfig{file=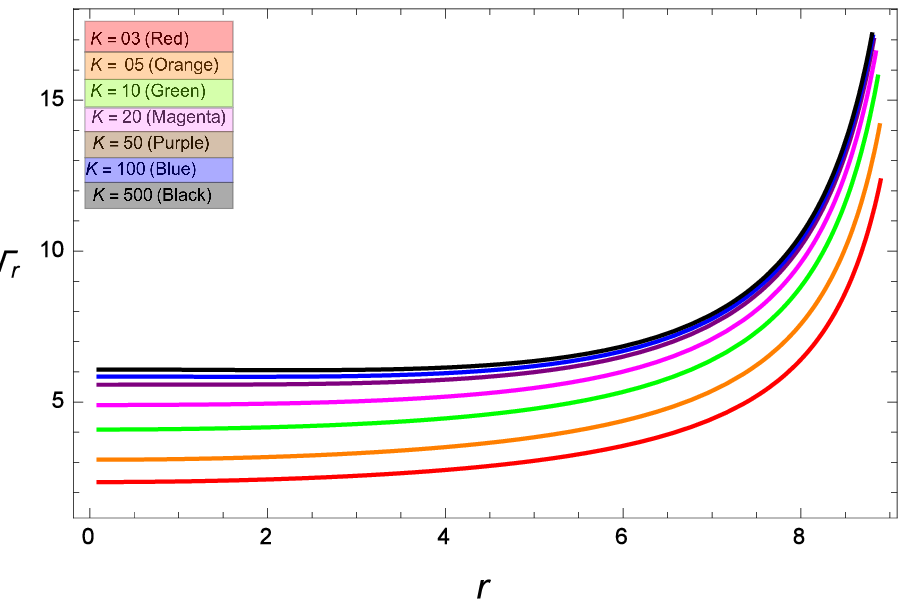, width=.48\linewidth,
height=2.3in}\epsfig{file=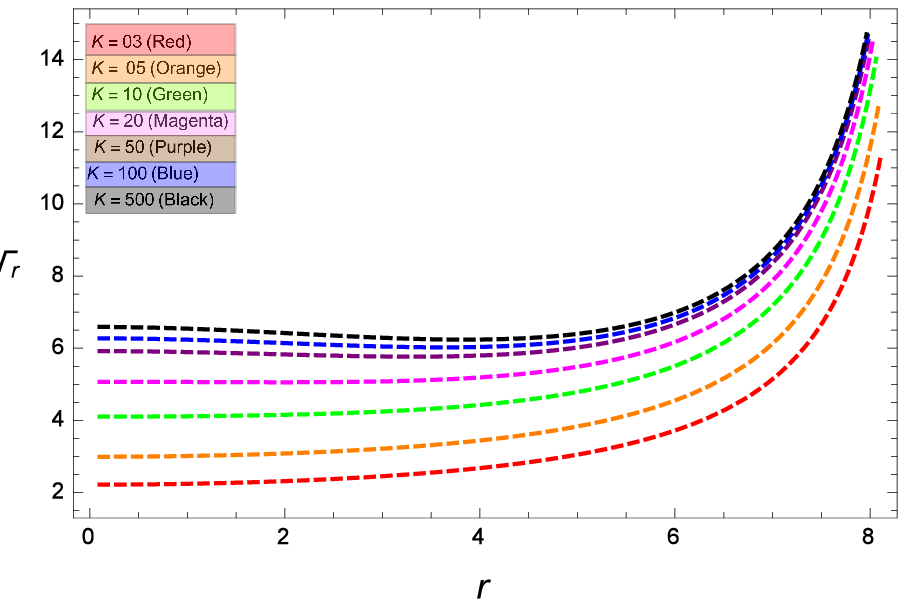, width=.48\linewidth,
height=2.3in}\caption{\label{Fig.6} Behavior of $\Gamma_{r}$.}
\end{figure}
From Fig. \textbf{19}, the graphical behavior of $\Gamma_{r}$ is shown monotonically increased and greater than $\frac{4}{3}$ for both models in $f(\mathscr{R},\mathscr{G})$ gravity with different values of parameter $K$.

\section{Conclusion}

In the current study, we explore the stellar structures in $f(\mathscr{R},\mathscr{G})$ gravity, with spherically symmetric space-time and anisotropic source of fluid. We aim to explore the two different compact star models namely LMC X-4, and EXO 1785-248 by employing the Karmarkar condition, which is used to embed the spherically symmetric space-time to class-1 metric. To our best of knowledge, this is the first attempt to investigate the stellar structures in the background of Karmarkar condition in $f(\mathscr{R},\mathscr{G})$ gravity. In this context, we have used the modified gravity model $f(\mathscr{R},\mathscr{G})=\mathscr{R}+\lambda \times \mathscr{R}^{2}+\mathscr{G}^2$ and calculated the corresponding field equations using the Karmarkar condition. Further, we use the matching condition, i.e., exterior Schwarzschild metric with interior space-time metric at the boundary to calculate the different values of involved parameters. The parameter $K$ involved in metric coefficients modeling becomes important for the analysis and we choose $K=3,5,10,20,50,100,500$. The estimated values of different parameters can be revealed from Table-\textbf{I} for both the stellar models. It is worth mentioning  that in this study, we have calculated the values of all the involved parameter by imposing the matching condition, i.e., there is no free parameter. We have shown the graphical analysis of the physical parameters, i.e., $\rho$,$\;p_{r},\;p_{t},\; \triangle,\;$$d\rho/dr, \;dp_{r}/dr$ and $d p_{t}/dr$. We have also investigated the stability of the compacts stars by exploring the energy conditions, equation of state, generalized TOV equation, causality condition, and adiabatic index. Some predicted values are also reported in tabular form for $e^{a(r=0)}$, $e^{b(r=0)}$, $\rho_{R}$, $p_{r_{c}}$, $p_{t_{c}}$, $\rho_{c}$, and \;$p_{r_{c}}/\rho_{c}=p_{t_{c}}/\rho_{c}$. Some important features of the present study are itemized below:
\begin{itemize}
  \item  From Table-\textbf{II}, it is depicted that $g_{rr}\mid_{(r =0)}=1$ and $g_{tt}\mid_{(r =0)}\neq0$. This behavior of metric potentials suggests that the Karmarkar condition is physical viable to show the stability of compact stars. It is worthwhile to mention here that in current study the $g_{tt}$ component is varied against the different values of parameter $K$, while the $g_{rr}$ component remains fixed as equal to 1.
  \item The Fig. \textbf{2} shows the graphical behavior of $\rho$ for two compact star models in $f(\mathscr{R},\mathscr{G})$ gravity. It is shown positive throughout the configuration for all the different values of parameter $K$. The behavior of pressure components is shown in Figs. \textbf{3} and  \textbf{4}. The radial pressure remains positive for $r<R$. Further, the tangential pressure ia also positive for both the models.
  \item  The anisotropy function is shown positive with regularly increasing behavior for both the models as shown in Fig. \textbf{5}. It is evident that the anisotropy function is zero at center due to equal values of radial and tangential values at center then it monotonically increases towards boundary and becomes maximum at $r=R$. This trend in the values of anisotropy function shows that our calculated results satisfy the stability conditions for both the stellar configurations.
  \item The derivatives  of energy density function and pressure components are also important in the present study. For $0<r\leq R$, we must have  $\frac{d\rho}{dr}< 0,\;\frac{dp_{r}}{dr}< 0,\;\frac{dp_{t}}{dr}< 0$ and at origin $\frac{d\rho}{dr}\mid_{r=0}= 0,\;\frac{dp_{r}}{dr}\mid_{r=0}= 0,\;\frac{dp_{t}}{dr}\mid_{r=0}= 0$. The negative nature of these gradients and behavior at origin can be seen from Figs. (\textbf{6}-\textbf{8}).
  \item The energy conditions $NEC$, $WEC$, $SEC$ and $DEC$ are seen satisfied for both the models in the current study. The graphical behavior can be revealed from Figs. (\textbf{9}-\textbf{11}). We have also calculated the two equation of state parameters $w_r$ and $w_t$. It is evident from Fig. \textbf{12} that both parameters remain positive and lie in the interval $(0,1)$.

  \item  The graphical behavior of gravitational red-shift function $Z_{s}$, compactness functions $u(r)$, and mass-radii function $m(r)$ is presented in Figs. (\textbf{13}-\textbf{15}). In current analysis, $Z_{s}\leq 2.400$, which is in good agreement with the range already predicted by Bohmer and Harko \cite{44} and Ivanov \cite{45}.  The compactness function $u(r)$ also satisfies the Buchdahl condition as in our case $u(r)\leq 0.30$. The mass-function $m(r)$ provides a monotonically increasing trend towards the surface.
  \item The balancing feature of anisotropic, hydrostatic, and gravitational forces can be seen from Fig. \textbf{16}. This balancing nature of $\mathscr{F}_{\mathrm{a}}$, $\mathscr{F}_{\mathrm{h}}$, and $\mathscr{F}_{\mathrm{g}}$ demonstrate that stellar structures under discussion are seen stable and physically acceptable.
  \item  It is seen in Fig. \textbf{17}, that both the radial and tangential velocities satisfy the condition $0\leq \nu_{r}~ \& ~\nu_{t}\leq1$. It is also evident in Fig. \textbf{18}, that the Abrea condition \cite{47}, i.e., $-1\leq \nu^{2}_{t}-\nu^{2}_{r}\leq 0$ is also fulfilled. Validity of both conditions confirms that our presented compact star models are potentially stable. The converse Abrea condition, i.e., $0\leq\nu^{2}_{r}-\nu^{2}_{t}\leq 1$, is also seen to be satisfied as shown in Fig. \textbf{18}. From Fig. \textbf{19}, the graphical behavior of $\Gamma_{r}$ is shown monotonically increased and greater than $\frac{4}{3}$ for both models in $f(\mathscr{R},\mathscr{G})$ gravity with different values of parameter $K$.
\end{itemize}
Hence, being sum-up, it is concluded that our obtained solutions are physically acceptable in the background of Karmarkar condition in $f(\mathscr{R},\mathscr{G})$ gravity. As a future work, it would be interesting to extend the analysis for other compact stars.\\\\
\section*{Appendix (\textbf{I})}
\begin{eqnarray*}
\lambda _1&&=\bigg(-2 F \left(L r^2+1\right)^K+K L r^2+K\bigg)\times\bigg(\frac{2 L^2 r^2 \chi _3(r) \left(F \left(L r^2+1\right)^K+2\right)}{\left(L r^2+1\right)^7 \chi _5(r){}^4}+\frac{L^5 r^8 \chi _3(r)}{\left(L r^2+1\right)^7 \chi _5(r){}^4}\\&&+\frac{L \chi _3(r)}{\left(L r^2+1\right)^7 \chi _5(r){}^4}+\frac{2 L^4 r^6 \chi _3(r) \left(F \left(L r^2+1\right)^K+2\right)}{\left(L r^2+1\right)^7 \chi _5(r){}^4}+\frac{L^3 r^4 \chi _3(r) \chi _{44}(r)}{\left(L r^2+1\right)^7 \chi _5(r){}^4}\bigg),\\
\lambda _2&&=\bigg(\frac{576 F^2 K^2 L^4 \chi _3(r){}^2 \left(L r^2+1\right)^{2 K-4}}{\chi _5(r){}^4 \chi _{35}(r){}^2}-\frac{L^5 r^8 \chi _{50}(r)}{\left(L r^2+1\right)^8 \chi _5(r){}^4}-\frac{L \chi _{50}(r)}{\left(L r^2+1\right)^8 \chi _5(r){}^4}\\&&-\frac{2 L^4 r^6 \chi _{50}(r) \left(F \left(L r^2+1\right)^K+2\right)}{\left(L r^2+1\right)^8 \chi _5(r){}^4}-\frac{2 L^2 r^2 \chi _{50}(r) \left(F \left(L r^2+1\right)^K+2\right)}{\left(L r^2+1\right)^8 \chi _5(r){}^4}-\frac{L^3 r^4 \chi _{44}(r) \chi _{50}(r)}{\left(L r^2+1\right)^8 \chi _5(r){}^4}\bigg),
\\
\lambda _3&&=384 F K L^4 \left(L r^2+1\right)^{K-6}\bigg(3 F \left(L r^2+1\right)^K-L r^2 \left(F \left(L r^2+1\right)^K-7\right)-K^2 L r^2 \chi _{37}(r)+K \chi _{36}(r)\\&&-3 L^3 r^6-L^2 r^4+5\bigg)\times\bigg(K \left(-L^2 r^4 \left(3 F^2 \left(L r^2+1\right)^{2 K}-2 F \left(L r^2+1\right)^K-12\right)-L^3 r^6 \left(3 F\right.\right.\\&&\times\left.\left. \left(L r^2+1\right)^K-8\right)+L r^2 \left(5 F \left(L r^2+1\right)^K+8\right)+2 L^4 r^8+2\right)+2 F^2 L r^2 \left(L r^2-1\right) \left(L r^2+1\right)^{2 K}\\&&+4 F K^2 L^2 r^4 \left(L r^2+1\right)^{K+1}\bigg),
\end{eqnarray*}
\begin{eqnarray*}
\lambda _4&&=\bigg(-2 F \left(L r^2+1\right)^K+K L r^2+K\bigg)\bigg(-\frac{12 L^2 \chi _3(r) \left(K-F \left(L r^2+1\right)^K\right)}{\left(L r^2+1\right)^7 \chi _5(r){}^4}-\frac{4 K (K+1) L^5 r^6 \chi _3(r)}{\left(L r^2+1\right)^7 \chi _5(r){}^4}\\&&-\frac{4 L^4 r^4 \chi _3(r) \chi _{43}(r)}{\left(L r^2+1\right)^7 \chi _5(r){}^4}-\frac{4 L^3 r^2 \chi _3(r) \chi _{45}(r)}{\left(L r^2+1\right)^7 \chi _5(r){}^4}\bigg),\\
\lambda _5&&=\frac{6 L^2 \chi _{50}(r) \left(K-F \left(L r^2+1\right)^K\right)}{\left(L r^2+1\right)^8 \chi _5(r){}^4}+\frac{2 K (K+1) L^5 r^6 \chi _{50}(r)}{\left(L r^2+1\right)^8 \chi _5(r){}^4}+\frac{2 L^4 r^4 \chi _{43}(r) \chi _{50}(r)}{\left(L r^2+1\right)^8 \chi _5(r){}^4}\\&&+\frac{2 L^3 r^2 \chi _{45}(r) \chi _{50}(r)}{\left(L r^2+1\right)^8 \chi _5(r){}^4},\\
\lambda _6&&=8 L^2 \chi _{38}(r)\bigg(K \left(-L r^2 \left(F^2 \left(L r^2+1\right)^{2 K}-12 F \left(L r^2+1\right)^K-16\right)-2 L^3 r^6 \left(F \left(L r^2+1\right)^K-4\right)\right.\\&&\left.+9 F \left(L r^2+1\right)^K+L^4 r^8+L^2 r^4 \chi _{41}(r)+5\right)+2 F K^3 L^2 r^4 \left(L r^2+1\right)^{K+1}-F \left(L r^2+1\right)^K \left(3 L^2 r^4\right.\\&&\left. \left(F \left(L r^2+1\right)^K+2\right)+5 \left(F \left(L r^2+1\right)^K+2\right)-2 L^3 r^6+L r^2 \chi _{40}(r)\right)+K^2 \left(L r^2+1\right) \chi _{39}(r)\bigg).
\end{eqnarray*}

\end{document}